\documentclass[11pt,oneside,letterpaper]{article}
\usepackage{amssymb}
\usepackage{amsmath}
\usepackage[dvips]{graphicx}
\usepackage{setspace}
\usepackage{fancyhdr}
\usepackage{xcolor}
\usepackage{ifpdf}
\usepackage{graphicx}
\usepackage{rotating}
\usepackage{comment}
\usepackage{color}
\usepackage{cite}
\definecolor{darkgreen}{rgb}{0,0.5,0}
\definecolor{darkblue}{rgb}{0,0,0.6}
\definecolor{purple}{rgb}{0.4,.2,0.7}
\usepackage[colorlinks=true,citecolor=darkgreen,linkcolor=purple,urlcolor=purple]{hyperref}

\newcommand{\reef}[1]{(\ref{#1})}

\newcommand{\co}{{\cal O}}

\newcommand{\be}{\begin{equation}}
\newcommand{\ee}{\end{equation}}

\def\bea{\begin{eqnarray}}
\def\eea{\end{eqnarray}}
\def\ba{\begin{array}}
\def\ea{\end{array}}
\def\bd{\begin{displaymath}}
\def\ed{\end{displaymath}}

\def\d{\delta}

\def\j{\psi}
\def\l{\lambda}
\def\m{\mu}
\def\n{\nu}
  
\def\s{\sigma}                                   

\def\D{\Delta}

\def\pa{\partial}                              
\def\>{\rangle} 
\def\<{\langle} 
\def\Dsl{D \hskip-.6em \raise1pt\hbox{$ / $ } }
\def\to{\rightarrow}
\def\pa{\partial}

\def\lab{\label}

\newcommand{\lra}{\leftrightarrow}

\def\vx{\vec{x}}
\def\vk{\vec{k}}

\def\l{\lambda}

\def\d{\delta}

\newcommand{\bi}{\begin{itemize}}
\newcommand{\ei}{\end{itemize}}

\numberwithin{equation}{section}
\allowdisplaybreaks  

\begin{document}
\begin{titlepage}

\begin{flushright}
{\tt MIT-CTP-4561}
\end{flushright}
\vspace*{2.3cm}

\begin{center}
{ \LARGE \textsc{Late-time Structure of the Bunch-Davies \\ De Sitter Wavefunction}\\}
\vspace*{1.7cm}
Dionysios Anninos$^1$, Tarek Anous$^2$, Daniel Z. Freedman$^{1,2,3}$ and George Konstantinidis$^1$

\vspace*{0.6cm}
$^1$ {\it Stanford Institute of Theoretical Physics, Stanford University} \\
$^2$ {\it Center for Theoretical Physics, Massachusetts Institute of Technology}\\
$^3$ {\it Department of Mathematics, Massachusetts Institute of Technology}

\vspace*{0.6cm}

danninos@stanford.edu, tanous@mit.edu, dzf@math.mit.edu, cgcoss@stanford.edu


\end{center}
\vspace*{1.5cm}
\begin{abstract}
\noindent

We examine the late time behavior of the Bunch-Davies wavefunction for interacting light fields in a de Sitter background.  We use  perturbative techniques developed in the framework of AdS/CFT,  and analytically  continue to compute tree  and loop level contributions to the Bunch-Davies wavefunction. We consider self-interacting scalars of general mass, but focus especially on the massless and conformally coupled cases.  We show that certain contributions  grow logarithmically in conformal time both at tree and loop level. 
We also consider gauge fields and gravitons. The four-dimensional Fefferman-Graham expansion of classical asymptotically de Sitter solutions is used to show that the wavefunction contains no logarithmic growth in  the pure graviton sector at tree level. Finally, assuming a holographic relation between the wavefunction and the partition function of a conformal field theory, we interpret the logarithmic growths in the language of conformal field theory.

\end{abstract}
\end{titlepage}

\newpage
\setcounter{page}{1}
\pagenumbering{arabic}

\tableofcontents
\setcounter{tocdepth}{2}

\onehalfspacing

\section{Introduction}

The geometry of the inflationary epoch of our early universe was approximately de Sitter \cite{Guth:1980zm,Linde:1981mu,Kazanas:1980tx,Starobinsky:1980te,Albrecht:1982wi}, and our universe is currently entering a de Sitter phase once again. It is thus of physical relevance to examine how to deal with quantum effects in a de Sitter universe. 
Such issues have been studied heavily in the past. The technical aspects of most calculations have involved the in-in/Schwinger-Keldysh formalism which is reviewed in \cite{Weinberg:2005vy}, and focus on computing field correlations at a fixed time. Indeed, in the context of quantum cosmology we are interested in correlations of quantum fields at a given time rather than scattering amplitudes---which condition on events both in the far past as well as in the far future.

A complementary approach is to build a perturbation theory for solutions of the Schr\"{o}dinger equation itself. Knowledge of the wavefunction allows us to consider expectation values of a broad collection of observables, which in turn permits a richer characterization of the state \cite{Anninos:2011kh}. Thus, an understanding of the wavefunction and its time evolution is of interest. Although generally complicated, there is one particular solution of the Schr\"{o}dinger equation in a fixed de Sitter background which exhibits a simplifying structure. This solution is the Bunch-Davies/Hartle-Hawking wavefunction $\Psi_{BD}$ \cite{Hartle:1983ai,Hertog:2011ky,Bunch:1978yq,Chernikov:1968zm}, and its form strongly resembles that of the partition function in a Euclidean AdS background upon analytic continuation of the de Sitter length and conformal time. This observation led to the elegant proposal of a close connection between dS and Euclidean AdS perturbation theory in \cite{Maldacena:2002vr} (see also \cite{Harlow:2011ke,Mata:2012bx}).

It is our goal in this paper to exploit the connection between dS and AdS to develop a more systematic perturbative framework for the construction of this wavefunction. We do this by considering a series of examples. The perturbative framework in an AdS spacetime has been extensively studied in the past \cite{Freedman:1998tz,D'Hoker:1998mz,D'Hoker:1999ni} and is our primary calculational tool. Our examples involve self-interacting scalar fields, both massless and massive, as well as gauge fields and gravitons. We recast many of the standard issues involving infrared effects of massless fields\footnote{See \cite{Weinberg:2006ac,Senatore:2009cf,Seery:2010kh,Freese:1986dd,Onemli:2002hr,Youssef:2013by,Gautier:2013aoa,Starobinsky:1994bd,Marolf:2010zp,Sasaki:1992ux,Prokopec:2006ue,Polyakov:2007mm,Tsamis:1996qq,Ford:1984hs,Giddings:2011ze,Burgess:2009bs,Burgess:2010dd} for an incomplete list of references on the topic of infrared issues in de Sitter space.} in the language of the wavefunction. Many of these infrared effects exhibit correlations that grow logarithmically in the scale factor, as time proceeds and we display how such effects appear in the wavefunction itself. It is worth noting that though most calculations of $\Psi_{BD}$ involve taking a late time limit, our approach requires  no such limit and we construct $\Psi_{BD}$ perturbatively for any arbitrary time. For massless scalar fields, the finite time dependence of the wavefunction at tree level is captured by the exponential integral function $\text{Ei}(z)$, whose small argument behavior contains the logarithmic contributions.

An interesting difference between the approach described in this paper and the in-in formalism is that the two approaches use different propagators. For a massless scalar in Euclidean AdS$_4$, we use the Green's function:\footnote{The Euclidean AdS metric is $ds^2 = L^2(dz^2+d\vec{x}^2)/z^2$ and we work in momentum space.}
\begin{multline} \lab{ads4}
G_{\text{AdS}}(z,z';k) = -\frac{1}{2 k^3 L^2} \bigg[(1-kz)(1+kz')e^{k(z-z')}\\ - 
\frac{e^{2 k {z_c}} (1-k {z_c})(1+kz)(1+kz')}{(1+k {z_c})}e^{-k(z+z')}\bigg],
\end{multline}
valid for $z<z'$. (For $z' < z$,  one simply exchanges the two variables.)
The mathematical purpose of the second term is to enforce the Dirichlet boundary condition at the cutoff $z_c$.  It is perhaps more significant physically  that the sum of the two terms is finite as  $k\to 0$.  Thus, loop integrals using (\ref{ads4})  do not produce infrared divergences at small $k$.  
The Green's functions considered in the in-in formalism \cite{Burgess:2009bs} are obtained by the continuation to dS$_4$ of the first term in square brackets.  Its real part gives 
\be
G_{C}(\eta,\eta';k) = \frac{1}{2k^3\ell^2}\bigg[(1+k^2\eta\eta')\cos[k(\eta-\eta')] + k(\eta-\eta')\sin[ k(\eta-\eta')]\bigg]~,
\ee
which is singular as $k \to 0$. 


One of the main motivations of our approach is to connect our results with the idea \cite{Maldacena:2002vr,Strominger:2001pn,Witten:2001kn} that $\Psi_{BD}$ (at late times) is holographically computed by the partition function of a conformal field theory. If this correspondence, known as the dS/CFT correspondence, is indeed true,\footnote{Recently several concrete realizations of this proposal have emerged \cite{Anninos:2011ui,Chang:2013afa,Anninos:2014hia} for theories of four-dimensional de Sitter space involving towers of interacting massless higher spin fields. Aspects of de Sitter holography are reviewed in \cite{Spradlin:2001pw,Anninos:2012qw}.}  infrared effects  in de Sitter spacetime should be related to quantities in the putative conformal field theory itself. This could lead to a better understanding of possible  non-perturbative  effects.  Moreover, in analogy with how the radial coordinate in AdS is related to some (as of yet elusive) cutoff scale in the dual CFT \cite{Heemskerk:2010hk,deBoer:1999xf,Kiritsis:2014kua,Mansfield:1999kk}, it is expected that the scale factor itself is connected to a cutoff scale in the CFT dual to de Sitter space \cite{Strominger:2001gp,Larsen:2002et,Bzowski:2012ih,Das:2013qea,Garriga:2013rpa}. Our calculations may help elucidate such a notion. Of further note, having a better understanding of $\Psi_{BD}$ at finite times allows us to compute quantum expectation values of fields within a single cosmological horizon, rather than metaobservables inaccessible to physical detectors.\footnote{A complementary approach would be to compute quantities directly in the static patch of de Sitter \cite{Anninos:2011af,Bernar:2014lna}.}

We begin in section {\ref{schroeqn}} by explaining how solutions to the Schr\"{o}dinger equation can be captured by a Wick rotation to Euclidean time, hence establishing the connection between de Sitter and anti-de Sitter calculations. We then proceed in section \ref{phi4sec} to examine a self-interacting scalar field with $\phi^4$ interactions in a fixed four-dimensional de Sitter background, whose contributions to the wave function contain terms that depend logarithmically on the  conformal time $\eta$. In section \ref{gaugesec} we discuss the case of gauge fields and gravitons. We argue that, to all orders in the tree-level approximation, no logarithms are present for a pure Einstein theory with a positive cosmological constant. We discuss our results in the context holography in section \ref{holo}. Finally, in section \ref{2dsec} we go to two-dimensional de Sitter space in order to compute loop effects for a cubic self-interacting massless scalar. In appendix \ref{toy} we set up a quantum mechanical toy model where the mathematics of our calculations is exhibited in a simple context.  

\section{The Schr\"{o}dinger equation in a fixed de Sitter background}\label{schroeqn}

The main emphasis of this section is to show that our method perturbatively solves
the functional Schr\"odinger equation for a scalar field in the Bunch-Davies state.  We will first provide the exact solution for a free field,  and then show how the result can be obtained by continuation from Euclidean AdS as in \cite{Maldacena:2002vr}.  We then treat interactions perturbatively.

We use conformal coordinates for dS$_{(d+1)}$,
\begin{equation}
ds^2 = \frac{\ell^2}{\eta^2}\left( {-d\eta^2 + d\vec{x}^2}\right)~, \quad\quad \vec{x} \in \mathbb{R}^d~, \quad\quad \eta \in (-\infty,0)~.
\end{equation}
For simplicity we consider a self-interacting scalar but analogous equations will also hold for other types of fields. The action is:
\begin{equation}
S_L = \frac{\ell^{(d-1)}}{2} \int_{\mathbb{R}^{d}} d \vec{x} \int \frac{d\eta}{|\eta|^{(d-1)}} \left( (\partial_\eta \phi(\eta,\vec{x}))^2 - (\partial_{\vec{x}} \phi(\eta,\vec{x}))^2 - \frac{\ell^2 \, V(\phi(\eta,\vec{x}))}{\eta^2} \right)~.
\end{equation}
We specify the potential later,  but we envisage the structure of a mass term
plus $\phi^n$ interactions.  

It is convenient to take advantage of the symmetries  of $\mathbb{R}^d$ and work in momentum space. Thus, we define:
\begin{equation}
\phi(\eta,\vec{x}) = \int_{\mathbb{R}^{d}} \frac{d\vec{k}}{(2\pi)^d} e^{i \vec{k} \cdot \vec{x}} \phi_{\vec{k}}(\eta)~.
\end{equation}
Henceforth we denote the magnitude of the  momentum by $k\equiv |\vec{k}|$. Upon defining the canonical momenta $\pi_{\vec{k}}= -i\delta/\delta \phi_{\vec{k}}$ conjugate to $\phi_{\vec{k}}$, we can write the Schr\"{o}dinger equation governing wavefunctions $\Psi[\varphi_{\vec{k}},\eta]$ in a fixed dS$_{d+1}$ background:
\begin{equation}  \lab{sch}
\sum_{\vec{k}\in\mathbb{R}^d}\left( \frac{1}{2}\frac{|\eta|^{(d-1)}}{\ell^{(d-1)}}\pi_{\vec{k}} \, \pi_{-\vec{k}} + \frac{\ell^{(d-1)}}{|\eta|^{(d-1)}}\left(\frac{ k^2}{2} \varphi_{\vec{k}} \, \varphi_{-\vec{k}} + \left(\frac{\ell}{|\eta|}\right)^2\tilde{V}(\varphi_{\vec{k}})\right)\right) \Psi[\varphi_{\vec{k}},\eta] = i \, \partial_\eta \Psi[\varphi_{\vec{k}},\eta]~.
\end{equation}
The variable $\varphi_{\vec{k}}$ is the momentum mode $\phi_{\vec{k}}$ evaluated at the time $\eta$ where $\Psi$ is evaluated. The potential $\tilde{V}(\phi_{\vec{k}})$ is the  Fourier transform of the original $V(\phi(\eta,\vec{x}))$; it has the structure of a convolution in $\vec k$-space. 

\subsection{Bunch-Davies wavefunction}

In principle, we can construct solutions to \reef{sch} by considering Feynman path integrals over the field $\phi$. 
We are particularly interested in the solution which obeys the Bunch-Davies boundary conditions. This state is defined by the the path integral:
\begin{equation}
\Psi_{BD} [\varphi_{\vec{k}},\eta_c] =  \int \prod_{\vec{k} \in \mathbb{R}^d} \mathcal{D} \phi_{\vec{k}} \,  e^{i S[\phi_{\vec{k}}]}~,
\end{equation}
in which we integrate over fields that satisfy $\phi_{\vec{k}} \sim e^{i k\eta}$ in the $k\eta \to - \infty$ limit and $\phi_{\vec{k}}(\eta_c) = \varphi_{\vec{k}}$ at some fixed time $\eta=\eta_c$. The natural generalization of this state to include fluctuating geometry at compact slicing is given by the Hartle-Hawking wavefunction. The boundary conditions resemble those defined in the path integral construction of the ground state of a harmonic oscillator. 

As usual, physical expectation values are given by integrating over the wavefunction squared. For example, the $n$-point function of $\varphi_{\vec{k}}$, all at coincident time $\eta_c$, is:
\begin{equation}\label{coscor}
\langle \varphi_{\vec{k}_1} \ldots  \varphi_{\vec{k}_n} \rangle =  \frac{\int  \prod_{\vec{k}} d\varphi_{\vec{k}} \, |\Psi_{BD}[\varphi_{\vec{k}},\eta_c] |^2 \, \varphi_{\vec{k}_1} \ldots  \varphi_{\vec{k}_n}}{ \int  \prod_{\vec{k}} d\varphi_{\vec{k}}\, |\Psi_{BD}[\varphi_{\vec{k}},\eta_c] |^2 }~.
\end{equation}

As a simple example we can consider the free massless field in a fixed  dS$_4$ background. In this case we can obtain $\Psi_{BD} $ as the exact solution of the Schr\"{o}dinger equation  \reef{sch}:
\begin{equation}\lab{psifree}
\Psi_{BD}[\varphi_{\vec{k}},\eta_c] = \prod_{\vec{k}}  \left({\frac{2k^{3}}{\pi }}\right)^{1/4} \exp\left[\frac{i\,\ell^2}{2}\left(\frac{k^2}{\eta_c(1 - i k \eta_c)}\right){\varphi_{\vec{k}}} \, {\varphi_{-\vec{k}}}\right]\frac{e^{-i k \eta_c/2 } }{\sqrt{(1-i k \eta_c)}} ~.
\end{equation}

Although $\eta_c$  can be considered to be an arbitrary point in the time evolution of the state,  we are ultimately interested in the late time structure of the wave function.
At late times, i.e. small negative $\eta_c$ we find:
\begin{equation}
\log\, \Psi_{BD}[\varphi_{\vec{k}},\eta_c] = \frac{\ell^2}{2} \, \int \frac{d\vec{k}}{(2\pi)^3} \, \left( \frac{i k^2}{\eta_c} - k^3 \right)\varphi_{\vec{k}} \, \varphi_{-\vec{k}} + \ldots~.
\end{equation}
Notice that the small $\eta_c$ divergence appears as a phase of the wavefunction rather than its absolute value, i.e. it plays no role in the expectation values of the field $\varphi_{\vec{k}}$. The late time expectation value of $\varphi_{\vec{k}} \, \varphi_{-\vec{k}}$ is given by:
\begin{equation}
\<\varphi_{\vec k} \, \varphi_{-\vec k}\> = \frac{1}{2 \, \ell^2 \, k^3}~,
\end{equation}
which diverges for small $k$. The divergence stems from the fact that $\Psi_{BD}$ is non-normalizable for the $\vec{k}=0$ mode. 


\subsection{A Euclidean AdS approach}\label{eucads}

When computing the ground state wavefunction of the harmonic oscillator from the path integral, one wick rotates time and considers a Euclidean path integral with boundary condition in the infinite past. Similarly, for the dS wavefunction, we can continue to Euclidean time $z = -i\eta$ and consider a Euclidean path integral.
Now, the path integral is over configurations that decay in the  infinite Euclidean past, defined here as  the limit $z \to \infty$. If in addition we continue $L = -i \ell $, we see that the calculation becomes that of constructing the Euclidean partition function in a fixed Euclidean AdS$_{(d+1)}$ background:
\begin{equation}\lab{adsbkgd}
ds^2 = \frac{L^2}{z^2}\left( {dz^2 + d\vec{x}^2}\right)~, \quad\quad \vec{x} \in \mathbb{R}^d~, \quad\quad z \in (0,\infty)~.
\end{equation}
In other words we have that: $\Psi_{BD}[\varphi_{\vec{k}},\eta_c] = Z_{AdS}[\varphi_{\vec{k}},i z_c]$ (with $L = -i \ell $) at least in the context of perturbation theory in a fixed (A)dS background.

The Euclidean path integral calculation incorporates, in principle at least, both classical and quantum effects.  Let us ignore quantum effects temporarily and discuss how AdS/CFT works at the classical level. To be concrete,  consider a massive scalar with quartic self-interaction. The classical action is 
\be \lab{adsact}
S = \frac{L^{(d-1)}}{2} \int_{\mathbb{R}^{d}} d \vec{x} \int_{z_c}^\infty \frac{dz}{z^{(d-1)}} \left( (\partial_z \phi)^2 +(\partial_{\vec{x}} \phi)^2 + \frac{m^2 L^2}{z^2} \phi^2 + \frac{\l \, L^2}{4 z^2} \phi^4  \right)~.
\end{equation}
One seeks a solution $\phi(z,\vec x)$ of the classical equation of motion that satisfies the boundary condition $\phi(z,\vec x)  \to \varphi(\vec x)$ as $z \to z_c$.  The cutoff $z_c$ is needed to obtain correct results for correlation functions in the dual CFT.  The classical solution is then substituted back in the action to form the on-shell action $S_{\text cl}[\varphi(\vec x)]$  which is a functional of the boundary data.  In the classical approximation the partition function is the exponential of the on-shell action i.e.   $Z_{\text AdS} = e^{-S_{\text cl}[\varphi(\vec x)]},$ and $n$-point correlation functions of the CFT operator dual to the bulk field $\phi$ are obtained by taking $n$ variational derivatives with respect to the sources $\varphi(\vec x)$.

Let us now perform the Euclidean version of the calculation that gives the result \reef{psifree}. For this purpose we ignore the quartic term in \reef{adsact}.
In $\vec k$-space,  we wish to solve the previously mentioned boundary value problem\footnote{There are many useful reviews of the AdS/CFT correspondence, including \cite{Aharony:1999ti,D'Hoker:2002aw, Nastase:2007kj}.  The present boundary value problem is discussed in Sec. 23.10 of \cite{Freedman:2012zz}.} captured by  the classical equation of motion:
\be\lab{adseom}
\bigg[ z^2\pa_z^2 - (d-1) z\pa_z - (k^2 z^2 + m^2L^2)\bigg] \phi_{\vec k}(z)=0\,,\qquad\quad \phi_{\vec k}(z=z_c) = \varphi _{\vec k}\,.
\ee
The exponentially damped solution of the ODE involves the modified Bessel function $K_\n(kz)$, and the solution of the boundary value problem can be neatly written as
\be \lab{btbdy}
\phi_{\vec k}(z) \equiv K(z;k) \varphi _{\vec k} = \frac{z^{d/2}K_\n(kz)}{z_c^{d/2}K_\n(k z_c)} \varphi _{\vec k}\qquad \quad \n=\frac12\sqrt{d^2+4m^2L^2}\,.
\ee
This equation defines the important bulk-to-boundary propagator $K(z,k)$.  

We follow the procedure outlined above and substitute the solution \reef{btbdy} into the action \reef{adsact}.  After partial integration the on-shell action reduces to the surface term at $z=z_c$:
\be \lab{scl2}
S_{ cl}[\varphi _{\vec k}] = -\frac{1}{2} \int \frac{d\vec k}{(2\pi)^d} \left(\frac{L}{z}\right)^{(d-1)}  K(z;k) \, \pa_z K(z;k) \, \varphi_{\vec k} \, \varphi _{-\vec k} \qquad {\text at }~~ z=z_c\,.
\ee
Let's restrict to the case of a massless scalar in AdS$_4$ which is the case $d=3,~\n = 3/2$ of the discussion above.  The Bessel function simplifies greatly for half-odd-integer index, and the bulk-to-boundary propagator becomes:
\be \lab{btbdymassless}
K(z;k) =\frac{(1+kz)e^{-kz}}{ (1+kz_c)e^{-k z_c}}\,.
\ee
The on-shell action then becomes:
\be \lab{s2D4}
S_{cl}[\varphi _{\vec k}]  = \frac{1}{2} \int \frac{d\vec k}{(2\pi)^3} \left(\frac{L}{z_c}\right)^2 \frac{k^2z_c}{1+kz_c} \varphi _{\vec k} \, \varphi _{-\vec k}\,.
\ee
To discuss the AdS/CFT interpretation we need to take the small $z_c$ limit, which gives:
\be
S_{{cl}} \to  \frac{1}{2}\int \frac{d\vec k}{(2\pi)^3} \, \left(\frac{L}{z_c}\right)^2 \, \left(k^2 z_c - k^3z_c^2 + \co(z_c^3)  \right)\varphi _{\vec k}\varphi _{-\vec k} \,.
\ee
The first term is singular as $z_c\to0$, but the factor $k^2 \varphi _{\vec k}\varphi _{-\vec k}$ is local in $\vec x$-space.  In fact it contributes a contact term $\d(\vec x-\vec y)$  in the $\vec x$-space correlation function. Such  contact terms are scheme-dependent in CFT calculations and normally not observable.  The remaining finite term has the non-local factor $k^3$.  It's Fourier transform
gives the observable part of the 2-point correlator,  $\<\co_3(x)\co_3(y)\>\sim 1/|\vec x -\vec y|^6$ which is the power law form of an operator of scale dimension $\D=3$. In AdS/CFT a bulk scalar of mass $m^2$ is dual to a scalar operator $\co_\D$ of conformal dimension $\D = (d/2 + \n)$. 

It is more pertinent to discuss the relation between the Lorentzian and Euclidean signature results.  In the free Lorentzian theory we can write $\Psi_{BD} = \exp (i S_{L})~$. Then upon continuation $L \to -i \ell,~z\to -i\eta, ~ z_c\to -i \eta_c$,   the Euclidean on-shell action \reef{s2D4}  and its Lorentzian counterpart are related by
\be \lab{EucLor}
-S_{E} \equiv - S_{cl} \to i S_{L} \,.
\ee
This is the expected relation for field theories related by Wick rotation.

Henceforth, the Euclidean signature AdS/CFT correspondence will be our primary method of computation.  In this way we will be using a well developed and well tested formalism.  After completion of a Euclidean computation, we will continue to de  Sitter space and interpret the results as contributions to the late time wave function $\Psi_{BD}$.

\subsection{Interaction corrections to $Z_{\text{AdS}} $}

We now consider the effect of interactions in the bulk action, such as, for example, the $\phi^4$ term in \reef{adsact}.  We treat the quantum fluctuations using a background field expansion $\phi = \phi_{\text cl}  + \d\phi$.
The classical field satisfies the non-linear classical equation of motion with Dirichlet boundary condition $\lim_{z\to z_c}\phi_{\text cl} (\vec{x},z) = \varphi(\vec{x})$, while the fluctuation $\d\phi$ vanishes at the cutoff.  The partition function is then:
\begin{equation}
Z_{\text{AdS}} [\varphi_{\vec{k}},z_c] = e^{-\, S_{cl}} \int \mathcal{D}\delta\phi \, e^{-S[\delta \phi,\phi_{cl}]}~.
\end{equation}
Exact solutions of the non-linear classical equation are beyond reach, but the reasonably efficient perturbative formalism of Witten diagrams  leads to series expansions in the coupling constant.  For example,   $\phi_{cl} = \phi_0 + \lambda \phi_1+\ldots$, where $\phi_0$ solves the free equation of motion coming from the quadratic piece of the action and the full Dirichlet boundary condition.\footnote{In Appendix A we present and develop a quantum mechanical toy model.  This model is instructive because perturbative computations are quite feasible  and their structure is closely analogous to those in our field theories.}

Witten diagrams without loops contribute to $S_{cl}$,  while those with internal loops appear in the perturbative development of the fluctuation path integral.  The basic building blocks of Witten diagrams are the bulk-to-bulk Green's function $G(z,w;k)$ and the bulk-to-boundary propagator $K(z;k)$.  It is significant that $G$ satisfies the Dirichlet condition at the cutoff, i.e. $G(z_c,w;k) =G(z,z_c;k) = 0$. 
In Appendix B, these propagators are explicitly constructed for the main
cases of interest in this paper. 

\begin{figure}[h!]
  \centering
    \includegraphics[width=0.5\textwidth]{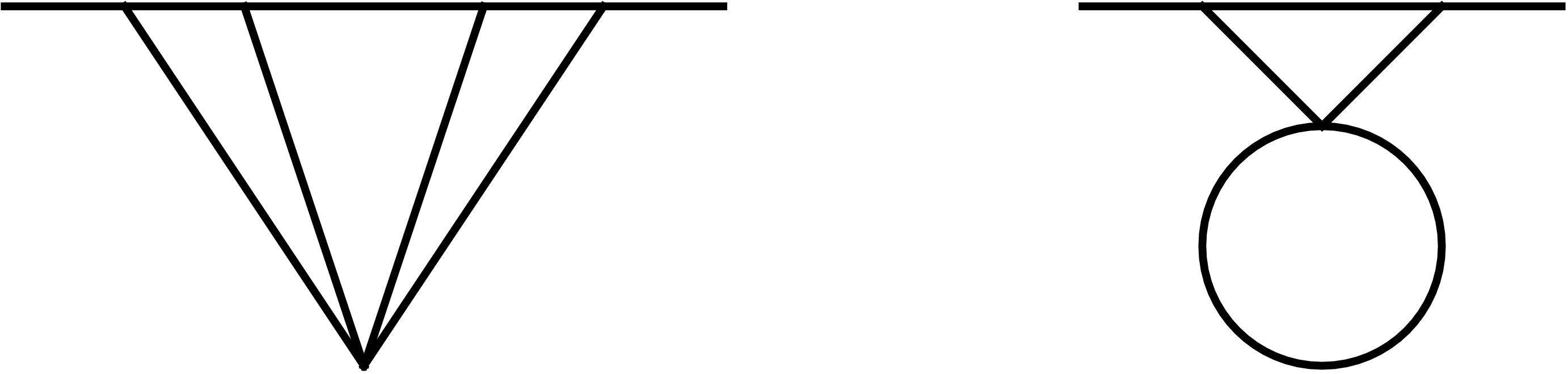}
    \caption{Witten diagrams for the order $\l$ contributions to $Z_{\text{AdS}} [\varphi_{\vec{k}},z_c]$ in the theory with $\phi^4$ interaction.}\label{phi4wittendiag}
\end{figure}

Witten diagrams for the order $\l$ contributions to $Z_{\text{AdS}} [\varphi_{\vec{k}},z_c]$ in the theory with $\phi^4$ interaction are depicted in 
Fig.~\ref{phi4wittendiag}.

\section{Self-interacting scalars in four-dimenions}\label{phi4sec}

We now discuss several contributions to $\Psi_{BD}$ from interactions, mostly in $\phi^4$ theory.  As previously mentioned, we carry out calculations in  Euclidean AdS$_4$, then continue to dS$_4$ by taking $z=-i\eta,~ z_c=-i \eta_c$ and $L = - i \ell$ .  We use the metric \reef{adsbkgd} and action \reef{adsact} in $d=3$.

\subsection{Tree level contributions for the massless theory}

First we focus on the massless case $m^2 L^2 = 0$. The relevant bulk-to-boundary propagator is given in \reef{btbdymassless}. The tree-level contribution to $\Psi_{BD}$ (left of Fig.~\ref{phi4wittendiag}) is captured by the integral:\footnote{The $\text{Ei}(z)$ function is defined as $\text{Ei}(z) = -\int_{-z}^\infty \, dt \, e^{-t}/t$. It has a branch cut along the positive real axis of $z \in \mathbb{C}$. We are primarily interested in this function along the negative real axis and the negative imaginary axis, both away from the origin.}
\begin{multline} \label{4ptlog} 
-\frac{\lambda \, L^4}{8} \, \int_{z_c}^\infty \prod_{i=1}^4 \frac{dz}{z^4} K(z;k_i) = \\ -\frac{\lambda L^4}{8} \, \frac{{k_\Sigma}+k_\Sigma^2 \, {z_c} + k_\Sigma \, \left(3 {k_\pi}-k_\Sigma^2\right) z_c^2+3 {k_P}\, z_c^3- \, e^{k_\Sigma z_c} {{k_\Sigma} \, {k_{\Sigma^3}} \, z_c^3 \, \text{Ei}(-k_\Sigma z_c)}}{3 \, k_\Sigma \, z_c^3 \, (1+{k_1} z_c) (1+{k_2} z_c) (1+{k_3} z_c) (1+{k_4} {z_c})}~,
\end{multline}
where we have defined the following quantities:
\begin{equation}\label{kdefs}
k_\Sigma \equiv \sum_{i=1}^4 k_i~, \quad k_\pi \equiv \sum_{\substack{i=1\\j=i+1}}^4 k_i k_j~, \quad k_P \equiv \prod_{i=1}^4 k_i~, \quad k_{\Sigma^3} \equiv  \sum_{i=1}^4 k_i^3~.
\end{equation}
We expand the result for small values of $z_c$ and analyze the divergent structure. In summary, we find terms of order $\sim z_c^{-3}$ as well as $\sim k^2 \, z_c^{-1}$ divergence but no $\sim z_c^{-2}$ term. Furthermore, we find a $\sim k^3 \log k \, z_c$ term. 

Upon analytic continuation to dS$_4$ the power law divergences become phases of the wavefunction. On the other hand, the logarithmic term contributes to the absolute value of $\Psi_{BD}[\varphi_{\vec{k}},\eta_c]$ at small $\eta_c$:
\begin{equation}\label{4pt}
\log|\Psi_{BD}[\varphi_{\vec{k}},\eta_c]| = \frac{\lambda \, \ell^4}{24} \, \int \frac{d\vec{k}_1}{(2\pi)^3}\, \frac{d\vec{k}_2}{(2\pi)^3} \,  \frac{d\vec{k}_3}{(2\pi)^3} \, \left( k_{\Sigma^3} \, \log(- k_\Sigma \, \eta_c) + \ldots \right) \, \varphi_{\vec{k}_1} \, \varphi_{\vec{k}_2} \,\varphi_{\vec{k}_3} \,\varphi_{\vec{k}_4}~,
\end{equation}
where $\sum_i \vec{k}_i = 0$ due to momentum conservation. Thus we encounter contributions to $|\Psi_{BD}[\varphi_{\vec{k}},\eta_c]|$ that grow logarithmically in the late time limit, $|\eta_c| \to 0$. In fact, at late enough times the correction is no longer a small contribution compared to the $\lambda=0$ pieces, and all subleading corrections will also begin to compete. In this way one recasts several of the infrared issues encountered when studying massless fields in the in-in/Schwinger-Keldysh formalism \cite{Weinberg:2005vy,Seery:2010kh}; now from the viewpoint of the wavefunction.

Similar logarithmic terms are present at tree level in a cubic self-interacting massless theory, and their effect was noted in the context of non-Gaussian contributions to inflationary correlators in \cite{Falk:1992sf}. In this case one finds the late time correction:
\begin{equation}\lab{3pt}
\log|\Psi_{BD}[\varphi_{\vec{k}},\eta_c]| = -\frac{\lambda \, \ell^4}{6} \, \int \frac{d\vec{k}_1}{(2\pi)^3}\, \frac{d\vec{k}_2}{(2\pi)^3} \left( k_{\Sigma^3} {\log}(- k_\Sigma \, {\eta_c}) + \ldots \right) \,  \varphi_{\vec{k}_1} \, \varphi_{\vec{k}_2} \,\varphi_{\vec{k}_3}~,
\end{equation}
where $\sum_i \vec{k}_i = 0$ , and $k_\Sigma$, $k_{\Sigma^3}$ are defined as in (\ref{kdefs}). In the case of slow roll inflation, these infrared effects are suppressed by the small slow roll parameters \cite{Maldacena:2002vr}.

As was mentioned in the introduction, the dS/CFT proposal connects $\Psi_{BD}$ to the partition function of a conformal field theory. Here, one envisions some theory in de Sitter space that contains such light scalars in its spectrum, including the graviton (dual to the stress tensor of the CFT) and so on. In section \ref{holo} we will explore this connection and in particular, discuss a possible holographic interpretation of such divergences based on recent analyses of 3d CFT's in momentum space \cite{Bzowski:2013sza,Coriano:2013jba}.

\subsection{Loop correction to the two-point function}\label{loop2}

It is of interest to understand the late time structure of loop corrections in the $\phi^4$ model. We will calculate the diagram on the right in Fig.~\ref{phi4wittendiag}, which corresponds  to the following integral:
\begin{equation}\label{loopint}
\mathcal{I}_{loop}(k,z_c) = - \frac{3 \, \lambda \, L^4}{4} \int_{z_c}^\infty \, \frac{dz}{z^4} \, K(z;k)^2 \, \int \frac{d\vec{p}}{(2\pi)^3} \,  G(z,z;p)~.
\end{equation}
To render the $\vec{p}$-integral finite we must impose an ultraviolet cutoff. Recall that $\vec{p}$ is a coordinate momentum, such that the physical (proper) momentum at a given $z$ is given by $\vec{p}_{ph} = z \, \vec{p}/L~$. We impose a hard cutoff on $\vec{p}_{ph}$, such that the ultraviolet cutoff of $\vec{p}$ is $z$-dependent, i.e. $|\vec{p}_{UV}| = |\Lambda_{UV}L|/z~$. A large $\Lambda_{UV}$ expansion reveals terms that diverge quadratically and logarithmcally in $\Lambda_{UV}$:\footnote{It is worth comparing the divergence structure in (\ref{div}) to a coincident point expansion of the $SO(4,1)$ invariant Green's function: $G(u) \sim \frac{1}{L^2}(2/u)^3 F(3,2,;4;-2/u)$.  The argument  $u = [(z-z')^2 + (\vec x-\vec x')^2]/2zz'$ is an  $SO(4,1)$ invariant variable. Near $u = 0$, we write  $z=z'$ and $\vec{x}=\vec{x}'+\vec{\epsilon}$.   In this limit $G(u) \sim \frac{1}{L^2}[- z^2/\epsilon^2 + 2 \ln(\epsilon/z)+ \ldots]~$. This is precisely of the form (\ref{div}), although  the divergence is cut off by  the physical length $\vec{x}_{ph,UV} = \vec{\epsilon} \, L/z$.  What we are suggesting is that the physical cutoff is a de Sitter invariant cutoff.}
\begin{equation}\label{div}
\frac{3 \, \lambda \, L^2}{8(2\pi)^2} \left(-{|\Lambda_{UV}L|^2} + 2 \log|\Lambda_{UV}L| \right)~.
\end{equation}
To cancel the quadratic divergence, we can add a local counterterm:
\begin{equation}
\delta \, \int \frac{dz}{z^4} \, \frac{d\vec{k}}{(2\pi)^3} \, \phi_{\vec{k}}(z) \, \phi_{-\vec{k}}(z)~, \quad\quad \delta = \frac{3 \, \lambda \, L^2}{8(2\pi)^2} \, {|\Lambda_{UV}L|^2}~.
\end{equation}
Upon addition of the counterterm, the $\vec{p}$-integral can be performed analytically rendering an expression containing the $\text{Ei}(z)$ function that is only logarithmically divergent in $|\Lambda_{UV}L|$. The remaining $z$-integral is complicated, but we are mainly interested in its small $z_c$ behavior, which we can extract. We find the following terms divergent in the small $z_c$ expansion (to leading order in $\Lambda_{UV})$:
\begin{equation}
-\frac{\lambda \, L^2}{2(2\pi)^2} \, {\log} |\Lambda_{UV} L| \, \left( -\frac{1}{{z_c}^3}+\frac{3 k^2}{{2 z_c}} + k^3 \, \log (z_c \, k) \right)~,
\end{equation} 
The logarithmic term contributes to the absolute value of the wavefunction upon analytic continuation to dS$_4$:
\begin{equation}
\log|\Psi_{BD}[\varphi_{\vec{k}},\eta_c]| = - \frac{\ell^2}{2} \, \int \frac{d\vec{k}}{(2\pi)^3} \, k^3 \, \left(1 - \frac{\lambda}{(2\pi)^2}  \log |\Lambda_{UV}\ell| \, \log(-\eta_c \, k) + \ldots \right)\varphi_{\vec{k}} \, \varphi_{-\vec{k}}~.
\end{equation}
Notice that at late times, the width of the $|\Psi_{BD}[\varphi_{\vec{k}},\eta_c]|$ for a fixed $k$ mode narrows, which is physically sensible as the quartic part of the potential dominates compared to the kinetic term.  To order $\lambda$, the ``cosmological two-point correlation function" can be obtained from this wave function (including the contribution from (\ref{4pt})) via the general expression (\ref{coscor}). 
The result closely resembles the late time two-point function computed, for example, in \cite{Burgess:2009bs}. Notice that there is no need to impose an infrared cutoff when considering loop corrections of the wavefunction itself.

As a final note, we could have also considered a slightly different subtraction where our counterterm also removes the logarithic divergence in $|\Lambda_{UV}L|$. Evaluation of the integrals proceeds in a similar fashion leading to the following result upon continuation to dS$_4$: 
\begin{equation}\label{shiftweight}
\log|\Psi_{BD}[\varphi_{\vec{k}}]| = - \frac{\ell^2}{2} \, \int \frac{d\vec{k}}{(2\pi)^3} \,  k^3 \left(1 + {a_1} \frac{\lambda}{(2\pi)^2}  \log (-\eta_c \, k)+ \ldots \right)\varphi_{\vec{k}} \, \varphi_{-\vec{k}}~,
\end{equation}
where $a_1 =  - (-5+4 {\gamma_E}+4\log2 )/4  \approx -0.02~$. The result is now independent of the ultraviolet cutoff altogether.

\subsection{Tree level contributions to the conformally coupled case}\label{conf}

We now analyze a conformally coupled scalar in a fixed Euclidean AdS$_4$ background with $m^2 L^2 = -2$. This case is of particular interest as it arises in the context of higher spin Vassiliev (anti)-de Sitter theories \cite{Vasiliev:1992av,Vasiliev:1995dn}. The bulk-to-boundary propagator simplifies to:
\begin{equation}\lab{skinconf}
K (z;k) = \left(\frac{z}{z_c}\right)\, e^{k(z_c-z)}~,
\end{equation}
and the free quadratic on-shell classical action is given by:
\begin{equation}
S_{cl} = \frac{L^2}{2} \, \int_{\mathbb{R}^3}  \frac{d\vec{k}}{(2\pi)^3} \, \varphi_{\vec{k}} \, \varphi_{-\vec{k}} \, \left( \frac{k}{z_c^2} - \frac{1}{z_c^3} \right)~.
\end{equation}
For the sake of generality, we consider a self-interaction of the form $\lambda_n \phi(\vec{x},z)^n / 2n~$ with $n=3,4,\ldots$ For such a theory, the order $\lambda_n$ tree level $(\varphi_{\vec{k}})^n$ contribution requires computing integrals of the form:
\begin{equation}
\mathcal{I}_n(k_i,z_c) = \int_{z_c}^\infty \frac{dz}{z^4} \, \left( \frac{z}{z_c} \right)^n e^{k_\Sigma (z_c-z)} = \frac{1}{z_c^3} \,e^{{k_\Sigma}\, {z_c}} {E_{(4-n)}}({k_\Sigma} \, {z_c})~,
\end{equation}
where $E_n(z)$ is the exponential integral function\footnote{The function $E_n(z) = \int_1^\infty dt e^{-z t}/t^n$ for $z\in \mathbb{C}$. It has a branch cut along the negative real axis. We are mostly interested in this function along the positive real axis and positive imaginary axis, both away from the origin.} and $k_\Sigma \equiv k_1+k_2+\ldots+k_n$. Expanding the integral reveals that logarithms will only occur in the small $z_c$ expansion for the case $n=3$. For $n=3$ we find the following small $z_c$ expansion:
\begin{multline}\label{3ptcs}
\mathcal{I}_{n=3}(k_i,z_c) = -\frac{\gamma_E+{\log}(k_\Sigma {z_c})}{{z_c}^3}-\frac{k_\Sigma (-1+\gamma_E + {\log}(k_\Sigma{z_c}))}{{z_c}^2} \\ -\frac{k_\Sigma^2 (-3+2 \gamma_E+2 {\log}(k_\Sigma {z_c}))}{4 {z_c}} - \frac{1}{36} k_\Sigma^3 (-11+6 \gamma_E+6 {\log}(k_\Sigma {z_c}))~.
\end{multline}
When we continue to dS$_4$ by taking $z_c = -i \eta_c$ and $L=-i\ell$, the leading contribution to the order $(\varphi_{\vec{k}})^3$ piece of the wavefunction at small $\eta_c$ is given by:
\begin{equation}
\log \Psi^{(3)}_{BD} =  -\frac{\lambda_3 \, \ell^4}{6} \, \int \frac{d\vec{k}_1}{(2\pi)^3} \frac{d\vec{k}_2}{(2\pi)^3} \, \varphi_{\vec{k}_1}\varphi_{\vec{k}_2}\varphi_{\vec{k}_3} \, \frac{1}{\eta_c^3} \, \left( -i \,{\gamma_E}  - i \log(-k_\Sigma \eta_c)+ \frac{\pi}{2} \right)~,
\end{equation}
with $\vec{k}_1 +\vec{k}_2 = -\vec{k}_3$ due to momentum conservation (see \cite{Boyanovsky:2011xn} for related calculations). We see that the absolute value of the wavefunction receives a $\sim 1/\eta_c^3$ divergent piece which is momentum independent (such that it becomes a contact term in position space). Interestingly, the cubic self-interaction of the conformally coupled scalar is absent in the classical Vasiliev equations \cite{Sezgin:2003pt,Petkou:2003zz}.

As another example, consider the quartic coupling which is conformal in four-dimensions. We find:
\begin{equation}
\mathcal{I}_{n=4}(k_i,z_c) = \frac{1}{z_c^4 \, k_\Sigma}~.
\end{equation}
Upon continuing to dS$_4$ this gives a momentum-dependent contribution to the real part of the exponent of the wavefunction, but none to the phase.

One can also consider loop corrections analogous to those computed in section \ref{loop2}. As an example we consider the one loop correction to the two-point function in the $\phi^4$ theory. The relevant Green function is given by:
\begin{equation}
G(z,w;k) = \frac{w z}{2kL^2} \, e^{-k(w+z)} \,\left( e^{2k z} - e^{2 k z_c} \right)~, \quad\quad z<w~,
\end{equation} 
and similarly for $z>w$. The relevant integral is (\ref{loopint}), though in this case there is only a quadratic divergence $\Lambda_{UV}^2$ to be cancelled. A small $z_c$ expansion of the regulated integral reveals the following contribution to the wavefunction:
\begin{equation}
\log |\Psi_{BD}[\varphi_{\vec{k}},\eta_c]| = - \frac{\ell^2}{2} \, \int \frac{d\vec{k}}{(2\pi)^3}\frac{k}{\eta_c^2} \left( 1 - \frac{3 \lambda_4}{4(2\pi)^2} \log(-k\eta_c)+\ldots \right) \varphi_{\vec{k}} \, \varphi_{-\vec{k}}~.
\end{equation}
Once again, we see that the wavefunction becomes narrower as time proceeds which is physically sensible.

\subsection{Comments for general massive fields}\label{massive4d}

We discuss the non-interacting case (with $\lambda = 0$) but non-zero mass. The solutions of the Klein-Gordon equation are given by:
\begin{equation}
\phi_{\vec{k}}(z) = \left(\frac{z}{z_c}\right)^{3/2} \frac{K_{\nu}( k z)}{K_{\nu}(k z_c)} \, \varphi_{\vec{k}}~, \quad \nu \equiv \sqrt{\frac{9}{4}+m^2 L^2}~.
\end{equation}
Once again we have imposed that the solution vanishes at $z\to \infty$. We are interested in the regime $\nu \in [0,3/2]$, since this range corresponds to light non-tachyonic scalars in dS$_4$ upon analytic continuation (such that $\nu = \sqrt{\frac{9}{4}-m^2 \ell^2}$). Heavy particles in dS$_4$ have pure imaginary $\nu$. The on-shell action is found to be:
\begin{equation}\label{mass2}
S_{cl} = - \frac{L^2}{2} \int \frac{d \vec{k}}{(2\pi)^3} \, \varphi_{\vec{k}} \, \varphi_{-\vec{k}} \, \left( \frac{(3-2 \nu)}{2\, {z_c^3}}-\frac{ k}{z_c^2 } \,\frac{ K_{(\nu-1)}( k {z_c})}{{ K_\nu}(k {z_c})} \right)~.
\end{equation}
For generic values of $\nu$ we can expand the action at small $z_c$ and find:
\begin{equation}
S_{cl} = - \frac{L^2}{2} \int \frac{d \vec{k}}{(2\pi)^3} \, \varphi_{\vec{k}} \, \varphi_{-\vec{k}} \, \frac{1}{z_c^3} \,\left( \frac{(3-2 \nu)}{2} - \frac{2\pi \, \text{csc}(\pi  \nu)}{ {\Gamma}(\nu)^2} \, \left(\frac{k z_c}{2}\right)^{2 \nu }   + \ldots \right)~.
\end{equation}
The above diverges at small $z_c$, in the region $\nu \in (0,3/2)$, even for the $\sim k^{2\nu}$ piece. In the context of AdS/CFT  the  boundary data for a scalar field with $\Delta-d = \nu -d/2 \neq 0$ must be ``renormalized" via $\varphi_{\vec{k}} \to z_c^{d-\Delta} \varphi_{\vec{k}}$ to achieve finite correlation functions as the cutoff $z_c\to 0$. (See Sec. 23.10 of \cite{Freedman:2012zz} for a discussion.) In the conformally coupled case discussed in section \ref{conf}, $\Delta = 2$ and $d=3$ such that $\varphi_{\vec{k}} \to z_c \, \varphi_{\vec{k}}~$. This renormalization absorbs the $1/z_c^2$ divergence. Upon continuing to dS$_4$ the $\sim 1/z_c^3$ term in (\ref{mass2}) becomes a phase and we find a factor $(i |\eta_c|)^{2(\nu-3/2)}$ which has a growing real part as $|\eta_c| \to 0$.

A small $z_c$ expansion in the $\nu=0$ case reveals $\sim \log k z_c$ terms in addition to the $1/z_c^3$ divergence. Only the logarithmic term contributes to the absolute value of $\Psi_{BD}[\varphi_{\vec{k}},\eta] $ upon continuing $z_c = -i \eta_c$ and $L = -i \ell$. In addition, we have that for $\nu=1$ there are also logarithmic terms in the small $z_c$ expansion. These become phases upon continuing $z_c = -i\eta_c$.

\subsubsection*{Tree-level diagrams}

Once again, we can ask whether the presence of logarithmic contributions to the late time wavefunction occur for more general values of $\nu$. The general bulk-to-boundary propagator is:
\begin{equation}
K (z;k) = \left(\frac{z}{z_c}\right)^{3/2} \frac{K_{\nu}( k z)}{K_{\nu}( k z_c)}~.
\end{equation}
Consider again self-interactions of the simple form $\lambda_n \phi(\vec{x},\eta)^n/2n$. The tree level integrals of interest are:
\begin{equation}\label{mtree}
\mathcal{I}^{(\nu)}_n(z_c,k_i) = \frac{L^4 \, \lambda_n}{2n} \, \int_{z_c}^\infty \,  \frac{dz}{z^4} \,\prod_{i=1}^n \, K(z;k_i)~.
\end{equation}
For generic $\nu$, we will find that at small $z_c$ the non-local piece in momentum will be accompanied by a divergent factor $z_c^{n(\nu-3/2)}~$. Upon continuation to dS$_4$, the local pieces which go as $1/z_c^3$ or $1/z_c$ will become phases of the wavefunction. On the other hand $z_c^{n(\nu-3/2)}$ will not contribute a pure phase to the wavefunction. However, upon computing a physical expectation value of $(\varphi_{\vec{k}})^n$ by integrating over the tree level $|\Psi_{BD}|^2$ one finds that it decays as $|\eta_c|^{n(3/2-\nu)}$ at late times. That the correlations decay in time for massive fields makes physical sense, since the particles dilute due to the expansion of space, and is consistent with a theorem of Weinberg \cite{Weinberg:2006ac}.

On the other hand, an examination of the small $z$ behavior of the Bessel $K_\nu(k z)$ function:
\begin{equation}
K_\nu(k z) = \frac{(k z)^{\nu}\, \Gamma(-\nu)}{2^{1+\nu}} \left( 1+ \frac{(k z)^2}{2(1+\nu)} + \ldots \right) +  \frac{(k z)^{-\nu}\, \Gamma(\nu)}{2^{1-\nu}} \left( 1+ \frac{(k z)^2}{2(1-\nu)} + \ldots \right)
\end{equation}
reveals that logarithmic terms can only occur of special values of $\nu$. They can only appear when the integrand of (\ref{mtree}) contains terms that go as $1/z$ in its small $z$ expansion, which integrate to a logarithm. For $n=3$, we have already discussed the massless $\nu = 3/2$ and conformally coupled $\nu = 1/2$ cases at tree level, as well as the $\nu = 0$ and $\nu=1$ cases at the free level. For general $n$, $\nu=3/2$ will still give rise to logarithmic contributions, as will $\nu = (3/2-3/n)$, where the logarithmic contributions are of the form $\sim\log(\sum_i k_i\, z_c)/z_c^{6/n}$ and $\nu = (3/2-1/n)$, where the logarithmic contributions are of the form $\sim\log(\sum_i k_i\, z_c)/z_c$ (in the range $\nu \in [0,3/2]$). In the latter case, the logarithmic contribution always appears as a phase upon analytic continuation to dS, not so for the former case.  It would be interesting if these are the only values of $\nu$ that give logarithms to higher order in perturbation theory. 

\section{Gauge fields and gravity in four-dimensions}\label{gaugesec}

In this section we consider classical contributions to the de Sitter wave function 
 for massless gauge fields and gravitons in a fixed dS$_4$ background. Compared with scalars treated in earlier sections, there are significant changes in the structure of the wave functions because of the gauge symmetry. 

\subsection{$SU(N)$ gauge fields}

In the non-Abelian case, the Yang-Mills action is given by:
\begin{equation}\label{ym}
S_{YM} = \frac{L^4}{4} \int_{\mathbb{R}^3} d \vec{x} \, \int_{z_c}^\infty \frac{dz}{z^4} \, \text{Tr} \,   F_{\mu \nu}^a F_{\rho\s}^a g^{\mu\rho}g^{\nu\s} ~, \quad\quad F^a_{\mu\nu}= \partial_{[\mu}A^a_{\nu]} + g \, f^{abc} A^b_\mu A^c_\nu~,
\end{equation}
where $a = 1,\ldots,N^2-1$ is the adjoint index and $f^{abc}$ are the $SU(N)$ structure constants. This action is conformally invariant at the classical level. 
This means that there will be no singular terms in $1/z_c$  in AdS vertex integrals and thus no terms in the de Sitter wave function that are logarithmically sensitive to $\eta_c$. 
The reason for this is that the  two inverse metrics in the action (\ref{ym}) soften the vertex integrals by a factor of $z^4$ and cancel the $1/z^4 $ from the metric determinant. 

We perform calculations in AdS$_4$ in the $A_z=0$ gauge, so only 
transverse spatial components of the gauge potential remain. 
In $\vec k$-space these components are given by
\bea
A_i (z,\vec k) &=& K(z;k) c_i(\vec k),  \qquad k_i c_i(\vec k) =0, \qquad c_i(\vec k)^* =c_i(-\vec k) \\
K(z;k) &=& e^{-k(z-z_c)}~.
\eea
The bulk-to-boundary propagator is so simple because $A_i$ obeys the same linearized equation as in flat space.  
Although usually not written explicitly,  the  transverse projector $\Pi_{ij} = \d_{ij} - k_ik_j/k^2$ is understood to be applied   to spatial vector modes. 

As our first calculation, we obtain the contribution of the free gauge field.
Metric factors cancel and we have the gauge-fixed action
\be \lab{sgf}
S = \int d\vec x \int_{z_c}^\infty \bigg[ \frac12 (\pa_z A_i)^2 + \frac14 (F_{ij})^2\bigg]\,.
\ee
After partial integration, as in  Sec.~\ref{eucads}, the on-shell action reduces to the surface term in $\vec k$-space:
\bea \lab{skingauge}
S &=& -\frac12 \int \frac{d\vec k}{(2\pi)^3} K(z,k)\pa_z K(z,k) c_i(\vec k) c_i(-\vec k)\\
&=& \frac12 \int \frac{d\vec k}{(2\pi)^3} k (\d_{ij} - k_ik_j/k^2)c_i(\vec k) c_j(-\vec k)\,.
\eea
The result contains the $\vec k$-space correlator of two conserved currents in the boundary 3d CFT.  This structure, which contains no dependence on $z_c$  may be compared with its analogue in \reef{skinconf} for the conformally coupled scalar.  The bulk fields $\phi$ and $A_i$ are both dual to CFT operators with $\D=2$.   There is only partial cancellation of metric factors for the scalar, so the singular factor $1/z_c^2$ remains.  As discussed in Sec.~\ref{massive4d},  this factor can be absorbed by renormalization of sources in AdS, but it gives a late-time power law singularity in $|\Psi_{BD}|$.

Next consider the tree-level three-point function.
The relevant integral is straightforward and gives a result with no dependence on $z_c$, namely:
\be \label{3ptym}
g \, f^{abc} \, \frac{ T_{ijk}}{k_1+k_2+k_3}\,
\ee
where $T_{ijk}$ is the same antisymmetric tensor that appears in flat space, namely:
\be \label{tensor}
T_{klm} = (\vec{k}_{1})_{l}\delta_{km} -(\vec{k}_{1})_{m}\delta_{kl} +(\vec{k}_{2})_{m}\delta_{kl} - (\vec{k}_{2})_{k}\delta_{lm} +(\vec{k}_{3})_{k}\delta_{lm}-(\vec{k}_{3})_{l}\delta_{km}\,.
\ee
Comparing (\ref{3ptym}) to the three point function of the conformally coupled scalar in (\ref{3ptcs}) we note the absence of logarithmic terms depending on $z_c$.

\subsection{Scalar QED}

Consider now a massive charged scalar field coupled to a $U(1)$ gauge field, with Euclidean action:
\begin{equation}\label{SphiA}
S_{SQED} = L^4 \, \int_{\mathbb{R}^3} d\vec{x} \int_{z_c}^\infty \frac{dz}{z^4} \left[ g^{\alpha\beta}(\partial_\alpha +i A_\alpha)\phi (\partial_\beta -i A_\beta)\phi^*  + m^2 \phi \, \phi^*\right]~.
\end{equation}
Properties of this theory were also considered  in \cite{Prokopec:2006ue}. Transverse modes in $\vec{k}$-space thus have the cubic interaction:
\be \label{SintphiA}
S_{int} = L^2 \, \int \frac{d\vec{k}_1}{(2\pi)^3} \, \frac{d\vec{k}_2}{(2\pi)^3} \, \int_{z_c}^\infty \frac{dz}{z^2} \, {A}_i(z,\vec{k}_3)\, (\vec{k}_1-\vec{k}_2)_i \, \phi_{\vec{k}_1}(z) \, \phi_{\vec{k}_2}^*(z)~,
\ee
where momentum  conservation requires  $\vec{k}_3= -\vec{k}_1-\vec{k}_2~$. Again, a transverse projector is understood to be applied to spatial vector modes. 
Using this interaction vertex, we find the following contribution to the partition function:
\be \label{amplit}
L^2 \int \frac{d\vec{k}_1}{(2\pi)^3} \, \frac{d\vec{k}_2}{(2\pi)^3} \, {\varphi}_{\vec{k}_1} \varphi_{\vec{k}_2} \,
{\tilde A}_i(\vec{k}_3)\, (\vec{k}_1-\vec{k}_2)_i \, \mathcal{I}_{m^2L^2}~, \quad\quad \tilde{A}_i(\vec{k}) \equiv A_i(z_c,\vec{k})~.
\ee
For a scalar field of mass $m^2 L^2$ and bulk-to-boundary propagator $K(z;k)$ the radial integral is
\be
\mathcal{I}_{m^2 L^2} =  \int_{z_c}^\infty \frac{dz}{z^2} \, e^{-k(z-z_c)} K(z;k_1)K(z;k_2) \,.
\ee
We compare the two cases of massless and conformally coupled $(m^2 L^2=-2)$ scalars with bulk-to-boundary propagators:
\be 
K_{m^2 L^2=0}(z;k)= \frac{(1+kz)e^{-kz}}{(1+kz_c)e^{-k z_c}}~, \quad\quad K_{m^2 L^2=-2}(z;k) =\left(\frac{z}{z_c}\right) e^{-k(z - z_c)}\,.
\ee
Our motivation is to explore the appearance of log$(kz_c)$ terms in the 3-point function. For the massless case we find:
\be
\mathcal{I}_{m^2 L^2=0} =
 \frac{\frac{1}{z_c}+ \frac{k_1 k_2}{k_1+k_2+k_3}+e^{(k_1+k_2+k_3) z_c} k_3 \text{Ei}[-(k_1+k_2+k_3) z_c]}{(1+k_1 z_c) (1+k_2 z_c)}
\ee
whose series expansion reveals a logarithmic term from the $\text{Ei}(z)$-function. 
%
For $m^2 L^2=-2$,  the integral is elementary and gives:
\be\label{confcase}
\mathcal{I}_{m^2 L^2=-2}  = \int_{z_c}^\infty dz \frac{e^{-(k_1+k_2+k_3)(z-z_c)}}{z_c^2}=\frac{1}{(k_1+k_2+k_3)z_c^2}~.
\ee
As in the case of the conformally coupled self-interacting scalar, we can absorb the $1/z_c^2$ divergence into a renormalization of the boundary data. 

After all is said and done, we find a $\sim \log(kz_c)$ term in the 3-point function of the massless scalar but not in the conformally coupled case. The ``practical" reason for the absence is the cancellation of the $\sim 1/z$ factors in the integrand of (\ref{confcase}) due to the softer behavior of the scalar bulk-to-boundary propagators. It would  be interesting to study the loop corrections to the wave function for these theories.
\subsection{Gravity}\label{treegr} 

It is a well known result that classical solutions in pure Einstein gravity with a positive cosmological constant $\Lambda = +3/\ell^2$  have a uniform late time (small $\eta$) expansion. In four spacetime dimensions, this is given by \cite{Starobinsky:1982mr}:
\begin{equation}\label{FG}
\frac{ds^2}{\ell^2} = -\frac{d\eta^2}{\eta^2} + \frac{1}{\eta^2} \left( g^{(0)}_{ij} + \eta^2 g^{(2)}_{ij} + \eta^3 g^{(3)}_{ij} + \ldots \right) dx^i dx^j~, \quad\quad |\eta|\ll 1~.
\end{equation}
The independent data in this expansion is the conformal class $\left(g^{(0)}_{ij},g^{(3)}_{ij}\right) \sim e^{\omega(\vec{x})}\,\left(g^{(0)}_{ij},g^{(3)}_{ij}\right)$. The Einstein equations  impose that $g^{(3)}_{ij}$ is transverse and traceless with respect to the boundary three-metric $g^{(0)}_{ij}$. Two of the phase space degrees of freedom reside in $g^{(0)}_{ij}$ and the other two in $g^{(3)}_{ij}$. The Einstein equations also require that the term linear in $\eta$ inside the parenthesis is absent. If $g^{(0)}$ and $g^{(3)}$ are appropriately related, the above solution will obey the Bunch-Davies boundary condition (this will require $g^{(3)}_{ij}$ to be complex). 
 
If $\Lambda<0$ there is an analogous expansions of the same structure known as the Fefferman-Graham expansion \cite{fg}. The on-shell action for such solutions satisfying the Bunch-Davies boundary condition (i.e. that the three-metric vanishes at large $z$ in EAdS) has been studied extensively \cite{Skenderis:2002wp}. Indeed, the on-shell  classical action is given at some fixed $z=z_c$ by:
\begin{equation}\label{FGS}
S_{gr} = \frac{3}{8\pi G L^2} \, \int_{\mathcal{M}} d\vec{x}  \int_{z_c}^\infty dz \sqrt{g} - \frac{1}{8\pi G L^2} \int_{\partial\mathcal{M}}d\vec{x} \sqrt{h} K^i_i~,
\end{equation}
where $h_{ij}$ is the induced metric on the fixed $z_c$ slice and $K_{ij}$ is the extrinsic curvature,  
\begin{equation}
K_{ij} = \frac{1}{2}\mathcal{L}_{n^\alpha} g_{ij}(z,\vec{x})~, \quad\quad n^\alpha =(z,\vec{0})~.
\end{equation}
The second term in (\ref{FG}), known as the Gibbons-Hawking term, is required for a well defined variational principle. For the first term we have used the on-shell condition $R = -12/L^2$. 

We can evaluate the classical action $(\ref{FGS})$ on the classical solutions obeying the Euclidean AdS$_4$ analogue of $(\ref{FG})$:
\begin{equation}\label{FGL}
\frac{ds^2}{L^2} = \frac{dz^2}{z^2} + \frac{1}{z^2} \left( g^{(0)}_{ij} + z^2 g^{(2)}_{ij} + z^3 g^{(3)}_{ij} + \ldots \right) dx^i dx^j~, \quad\quad z \ll 1~.
\end{equation}
 and expand in small $z_c$.
The expansion of the on-shell classical action contains only divergences of the form ${1}/{z_c^3}$ and ${1}/{z_c}$ at small $z_c$, but no $1/z_c^2$ or $\log z_c$ divergences \cite{Henningson:1998gx}. The absence of a $\sim 1/z$ term in (\ref{FGL}) is crucial for the logs to be absent in the small $z_c$ expansion of the on-shell classical action.\footnote{Note that in odd space-time dimensions, there is a piece 
of the Fefferman-Graham expansion which contributes logarithmic terms to the phase of the wavefunction as well as local terms to its absolute value \cite{Maldacena:2002vr}.} The divergent terms amount to pure phases in $\Psi_{BD}[g_{ij},\eta]$ upon analytic continuation.\footnote{There is a slight subtlety in assuming that the full solution $ds^2/\ell^2 = -d\eta^2/\eta^2 + g_{ij}(\vec{x},\eta) dx^i dx^j$ allowing for the expansion (\ref{FG}) can indeed by analytically continued to $z = -i\eta$ at the non-linear level. For small enough deviations away from the flat metric $g^{(0)}_{ij} = \delta_{ij}$  the bulk-to-bulk and bulk-to-boundary propagators allow for such a continuation.} (See \cite{Pimentel:2013gza} for a related discussion.) The important point is that there are no logarithmic divergences for small $z_c$, which translates to the statement that the Bunch-Davies wavefunction exhibits no $\sim k^3 \log (-\eta_c \, k)$ growth at tree level. 

Thus the Fefferman-Graham expansion for dS$_4$ seems to explain the absence of logarithms in the gravitational 3-point functions calculated, for example, in \cite{Maldacena:2011nz,Bzowski:2011ab}. This is in stark contrast to the case of the massless scalar.


\section{3d CFTs and (A)dS/CFT}\label{holo}

The dS/CFT correspondence proposes that the Bunch-Davies (or Hartle-Hawking) wavefunction of dS$_4$ at late times is computed by the partition function of a three-dimensional Euclidean conformal field theory. It is closely related to the Euclidean AdS/CFT proposal, as we have tried to make clear above. In the AdS/CFT context the small $z_c$ cutoff is identified with a cutoff in the dual theory. This is due to the manifestation of the dilatation symmetry as the $(z,\vec{x}) \to \lambda(z,\vec{x})$ isometry in the bulk. For instance, bulk terms that diverge as inverse powers of $z_c$ (with even powers of $k$) are interpreted as local terms in the dual theory. On the other hand, the tree level $z_c$-dependent logarithmic terms, such as those in the small $z_c$ expansion of (\ref{4ptlog}), are \emph{not} local in position space and yet seem to depend on the cutoff. One may ask whether they have an interpretation from the viewpoint of a putative CFT dual. 

Recent analyses of CFT correlation functions in momentum space \cite{Bzowski:2013sza,Coriano:2013jba} give a suggestive answer. Recall that the symmetries of CFTs have associated Ward identities, governing correlation functions. For concreteness we specifically consider the Ward identities, expressed in momentum space, constrainging the three point functions of a scalar operator $\mathcal{O}$ with weight $\Delta$. The Ward identity for the dilatation symmetry is given by:
\begin{equation}\label{ward}
\left( 6+ \sum_{i=1}^3 (p_j \partial_j - \Delta)  \right) \langle \mathcal{O}(\vec{p}_1)  \mathcal{O}(\vec{p}_2) \mathcal{O}(\vec{p}_3)  \mathcal \rangle = 0~,
\end{equation}
whereas for the special conformal transformations we have:
\begin{equation}
\sum_{i=1}^3 (\vec{p}_i)^\alpha \, \left( \partial^2_i + \frac{4 - 2\Delta}{p_i} \partial_i \right) \langle \mathcal{O}(\vec{p}_1)  \mathcal{O}(\vec{p}_2) \mathcal{O}(\vec{p}_3)  \mathcal \rangle = 0~.
\end{equation}
The Latin index labels a particular momentum insertion, $\mathcal{O}(\vec{p}_i)$, whereas the Greek index labels a specific Euclidean component of $\vec{p}_i$. We have also removed the $\delta(\vec{p}_1+\vec{p}_2+\vec{p}_3)$ conservation rule from the correlator.  The solution to the above equations is most conveniently expressed as an integral over an auxiliary coordinate:
\begin{multline}\label{cftintegral}
\langle \mathcal{O}(\vec{p}_1)  \mathcal{O}(\vec{p}_2) \mathcal{O}(\vec{p}_3)  \mathcal \rangle = \\ c_3 \left( p_1 p_2 p_3 \right)^{\Delta-3/2} \int_0^{\infty} dz\, z^{1/2} K_{\Delta-3/2}(z \, p_1)\,K_{\Delta-3/2}(z \, p_2)\,K_{\Delta-3/2}(z \,p_3)~,
\end{multline}
where $K_\nu(z)$ is the modified Bessel function of the second kind. The above integral should look familiar; we have indeed encountered it in our previous analysis of massive fields.
As previously noted, from the bulk point of view the conformal weight of a scalar of mass $m^2 L^2$ in AdS$_4$ is $\Delta = 3/2+\nu$ where $\nu \equiv \sqrt{9/4+m^2 L^2}$. The case $\nu = 3/2$, i.e. a massless scalar field in AdS$_4$, corresponds to a marginal operator with $\Delta = 3$. 
Thus, the auxiliary variable $z$ can be precisely identified with an AdS bulk coordinate, and the modified Bessel functions can be thought of as bulk-to-boundary propagators (\ref{btbdy}). 
The integral (\ref{cftintegral}) is of course divergent for general $\Delta$ near $z=0$. Motivated by our bulk analysis, we chose a slightly different cutoff procedure from \cite{Bzowski:2013sza}, where we instead cut the integral off at some small $z=z_c$. 
Now from a CFT analysis, we find the appearance of logarithmic contributions, which will generally be cutoff dependent, to the three-point function of a scalar operator (this observation remains true even in the cutoff prescription chosen in  \cite{Bzowski:2013sza}). 
Because these logarithmic terms contain a dependence on the cutoff scale $z_c$, they are consequently referred to as anomalies in \cite{Bzowski:2013sza}.
They may be present in the theory non-perturbatively. Thus, from the holographic point of view, terms logarithmic in $z_c$ that are associated to anomalies in the 3d CFT, such as those in the three-point function, will be present to all orders rather than part of a resummeable series. 

Let us note that we also observed such logarithms in higher point functions at special values of $\nu$ (e.g. $\nu=3/2$), where the analogous general CFT analysis is more cumbersome. In a similar fashion, the tree level logarithms we discussed for the Bunch-Davies wavefunction in section \ref{massive4d} (for $\nu=0$ or $\nu=1$) are related to a divergence in the Fourier transform of the two-point function of a weight $\Delta = 3/2$ or $\Delta= 5/2$ scalar operator \cite{Osborn:1993cr,Bzowski:2013sza}.

It is also possible, however, that the appearance of these logarithms are the result of small shifts in the conformal weights of certain operators in the 3d CFT. 
For instance, imagine that loop corrections (such as $1/N$ corrections in a large $N$ CFT) shift $\Delta^{(0)}$ by an order $\sim1/N$ amount, i.e. $\Delta = \Delta^{(0)} + \alpha/N + \mathcal{O}(1/N^2)$ with $\alpha \sim \mathcal{O}(1)$. The two point function in momentum space will then have a large $N$ expansion:
\begin{equation}
k^{2(\Delta-3/2)} = k^{2(\Delta^{(0)}-3/2)} \left( 1 + \frac{2\alpha}{N} \log k + \ldots \right)~. 
\end{equation}
From the bulk AdS$_4$ point of view, we must include the factors $z_c^{2(\Delta-3)}$ to obtain the $z_c$-dependent bulk partition function, as we discussed in section \ref{massive4d}, such that the expansion becomes:
\begin{equation}\label{weightcorrection}
\frac{(z_c k)^{2(\Delta-3/2)}}{z_c^3} = \frac{(z_c \, k)^{2(\Delta^{(0)}-3/2)}}{z_c^3} \left( 1 + \frac{2\alpha}{N} \log (z_c k) + \ldots \right)~. 
\end{equation}

As we already noted, we can extrapolate the perturbative results in AdS to those in dS by continuing $z_c = -i\eta_c$ and $L = -i\ell$. From the point of view of a putative dual CFT of dS, the $z_c=-i\eta_c$ continuation corresponds to an analytic continuation of the cutoff itself. 
Though unusual from the point of view of field theory, it may be interesting to consider general properties of field theories with such imaginary cutoffs. Notice that the expansion (\ref{weightcorrection}) now contains $\sim \log(-k\eta_c)$ pieces which are resummed to a power law behavior in $\eta_c$.\footnote{A concrete realization occurs in the conjectured duality between the three-dimensional $Sp(N)$ critical model \cite{Anninos:2011ui,Anninos:2013rza} and the minimal higher spin theory in dS$_4$. The bulk scalar has a classical mass $m^2\ell^2=+2$ and is dual to a spin zero operator whose conformal weight is $\Delta=2$ at $N=\infty$, but receives $1/N$ corrections \cite{brezin} (related by $N\to-N$ to those of the critical $O(N)$ model). Similar corrections will also occur for the extended dS/CFT proposals in \cite{Chang:2013afa,Anninos:2014hia}.} With this interpretation we might view (\ref{shiftweight}) as a small negative shift in the weight $\Delta = 3$ of the relevant operator dual to the bulk massless field, such that it becomes slightly relevant. On the other hand, the fact that the three-point function of a marginal scalar operator contains an anomalous logarithm suggests that the wavefunction has a non-trivial time evolution. In the case we consider, where it is due to a cubic self-interaction of a massless scalar (see \reef{3pt}), this might have been expected given that we are not perturbing about a stable minimum of the bulk scalar potential. However, this anomalous logarithm may disappear should we correct the propagators to reflect the negative shift in weight.  

The CFT stress tensor operator $T_{ij}$ also weight $\Delta = 3$ and is thus also a marginal operator. In (A)dS/CFT it is dual to the bulk graviton. Absence of a Weyl anomaly in three dimensional CFTs can be expressed as the following property of the CFT partition function:
\begin{equation}
Z_{CFT}[g_{ij}] = Z_{CFT}[e^{\omega(x)}g_{ij}]~,
\end{equation}
where $\omega(x)$ is a smooth function and we are removing local counterterms. The above implies that correlation functions of the stress tensor, given by variational derivatives with respect to $g_{ij}$, cannot depend on the Weyl factor of $g_{ij}$ (in the absence of any other sources) and in particular cannot depend on the logarithm of the cutoff. This strongly suggests, if we are to take the picture of dS/CFT seriously, that late time $\log \eta_c$ contributions to the wavefunction $\Psi_{BD}[g_{ij}]$, such as the one describing pure Einstein theory, are absent to all orders in perturbation theory. This agrees with several computations of the cubic contribution \cite{Maldacena:2011nz,Bzowski:2011ab}, as well as our general tree level argument in section \ref{treegr}, which are all devoid of such logarithmic terms. These observations, however, do not preclude the possibility of $\Psi_{BD}[g_{ij},\eta]$ peaking far from the de Sitter vacuum. 


\section{dS$_2$ via Euclidean AdS$_2$}\label{2dsec}
We now proceed to study several perturbative corrections of the Bunch-Davies wavefunction about a fixed dS$_2$ (planar) background.
We consider the massless  scalar field  in Euclidean AdS$_2$  whose action is:
\begin{equation}\label{cubic}
S_E = \frac{1}{2} \, \int_{\mathbb{R}} d\vec{x} \int_{z_c}^{\infty} {d z} \left( (\partial_z \phi(\vec{x},z))^2 + (\partial_{\vec{x}}\phi(\vec{x},z))^2 + \frac{L^2 \, m^2}{z^2} \phi(\vec{x},z)^2 + \frac{L^2 \, \lambda}{3 \, z^2} \, \phi(\vec{x},z)^3 \right)~.
\end{equation}
The reason for reducing to two-spacetime dimensions is that the integrals needed
for order $\lambda^2$ calculations are far simpler that for AdS$_4$, although their mathematical structure and the physical issues are quite similar. We will focus on cubic interactions.

\subsection{Tree level corrections for the massless theory}

The simplest contribution to consider is the order $\lambda$ $(\varphi_{\vec{k}})^3$ contribution. For  massless fields, the bulk-to-boundary $K(z;k)$ and bulk-to-boundary $G(z,w;k)$ propagators are given in appendix \ref{GK}.
This correction is a tree level diagram involving three bulk-to-boundary propagators. In order to calculate it, we must evaluate the integral:
\begin{equation}\label{cubicintegral}
-\frac{L^2 \, \lambda}{6} \int_{z_c}^\infty \frac{dz}{z^2} \, K(z;k_1) K(z;k_2) K(z;k_3) =- \frac{L^2 \, \lambda}{6} \left( \frac{1}{{z_c}}+e^{{z_c}\,k_\Sigma } \, k_\Sigma \, {\text{Ei}}(-{z_c} \, k_\Sigma ) \right)~,
\end{equation}
where $k_\Sigma \equiv k_1 + k_2 + k_3$. In the limit of small $z_c$ we find a $\sim \log(z_c \, k_\Sigma)$ contribution. Continuing  to dS$_2$ by taking $L = -i \ell$ and $z_c =  -i\eta_c$, the Bunch-Davies wavefunction at late times to order $\lambda$ is given by:
\begin{equation}
\log \Psi_{BD} = \int \frac{d \vec{k}_1}{2\pi} \left( - \frac{k}{2} \, \varphi_{\vec{k}_1}\varphi_{-\vec{k}_1} + \frac{\ell^2\lambda}{6}\int \frac{d\vec{k}_2}{2\pi} \varphi_{\vec{k}_1}\varphi_{\vec{k}_2} \varphi_{\vec{k}_3} \left( \frac{i}{
\eta_c} + k_\Sigma \big(\gamma_E + \log(-\eta_c k_\Sigma)\big)   \right)   \right)~,
\end{equation}
where we must impose $\vec{k}_3 = -\vec{k}_1 - \vec{k}_2$ due to momentum conservation.  Once again, we note that the absolute value of the Bunch-Davies wave function receives a logarithmic contribution.

At order $\lambda^2$ we have a $\sim (\varphi_{\vec{k}})^4$ contribution to the wave function which also involves an integration over the bulk-to-bulk propagator. The integrals can also performed to obtain a result that behaves (schematically) in the small $z_c$ limit as $\sim \lambda^2 k \log^2 k z_c~$. The integral we need is:
\begin{equation}\label{fpf}
\frac{\lambda^2 \, L^4 }{8}\int_{\mathcal{D} } \frac{dz}{z^2} \, \frac{dw}{w^2} \, G\left(z,w,\vec{q}\right)  \, K\big(z;{k_1}\big) \, K\big(z;{k_2}\big)  \, K\big(w;{k_3}\big) \, K\big(w;{k_4}\big)~,
\end{equation}
where the domain of integration is $\mathcal{D} = [z_c,\infty]^2$. In the small $z_c$ limit, we find:  
\begin{equation}\label{exactexp}
\frac{\lambda^2 \, L^4}{8}\bigg\lbrace\frac{1}{z_c}
-\frac{1}{2}(s+q)\left(\log[(s+q)z_c]\right)^2-\frac{1}{2}(t+q)\left(\log[(t+q)z_c]\right)^2 +\dots\bigg\rbrace~, 
\end{equation}
where $s\equiv |\vec{k}_1|+|\vec{k}_2|$ and $t\equiv |\vec{k}_3|+|\vec{k}_4|$ and $q\equiv |\vec{q}|=|\vec{k}_1+\vec{k}_2|=|\vec{k}_3+\vec{k}_4|$  (note that $s,t>q$ by the triangle inequality). 


\subsection{Loop corrections for the massless theory}

A tadpole diagram contributes to the wavefunction at order $\lambda$. The relevant integral is given by:
\begin{equation}\label{tadpole}
\frac{L^2 \, \lambda}{2} \, \int_{z_c}^\infty \frac{dz}{z^2} K(z,k=0)\, \int \frac{d\vec{p}}{2\pi} \, G(z,z;p)~.
\end{equation}
Note that $ K(z,k=0) \equiv 1$.
To render the integral finite we impose a physical ultraviolet cutoff, which becomes a $z$-dependent cutoff $p_{UV} = \Lambda_{UV} L/z$ for the coordinate momentum over which we are integrating. We can add a counterterm to the action of the form:
\begin{equation}
S_{ct} =  \delta \, L^2 \int_{\mathbb{R}} {d\vec{x}} \, \int^{\infty}_{z_c} \, \frac{d z}{z^2} \, \phi(\vec{x},z)~.
\end{equation}
The constant $\delta$ can be selected to cancel the logarithmic divergence in $\Lambda_{UV}$ rendering the following result for the full integral (\ref{tadpole}):
\begin{equation}
\frac{L^2 \, \lambda}{4\pi} \, \left( \frac{-1 + \gamma_E + \log 2}{z_c} \right)~.
\end{equation}
Upon continuation to dS$_2$ this contributes only to the phase of the wavefunction.  

At order $\lambda^2$ we have two distinct loop corrections to the $\sim (\varphi_{\vec{k}})^2$ term. One involves attaching a tadpole to the tree level propagator whose ultraviolet divergence can be treated as above. The relevant integral is given by:
\begin{equation}
\mathcal{I}_{tadpole}(k,z_c;L) = \frac{L^4 \, \lambda^2}{4} \, \int_{\mathcal{D} } \frac{dw}{w^2}\frac{dz}{z^2} \, \int \frac{d\vec{p}}{2\pi} \, G(w,w;p)  \, G(z,w;0) \, K(z;k) \; K(z;k)~.
\end{equation}
At small $z_c$ the above integral  contains a finite term plus  a logarithmic piece in $z_c$.  The result is:
\begin{equation}
\mathcal{I}_{tadpole}(k,z_c;L) = \frac{L^4 \, \lambda^2}{2\pi}   \left( \frac{-12+ \pi^2+6\left(\gamma_E+\log 2 \right)}{24{z_c}}- \frac{\left(\gamma_E+\log 2 \right)}{4} k \, ({\log}k {z_c})^2 + \ldots  \right)~. 
\end{equation}
where the subleading pieces are at most logarithmic in $z_c$.

The other order $\lambda^2$ contribution comes from a `sunset' diagram, which is ultraviolet finite in two-dimensions and thus requires no regularization. It involves an integral of the form:
\begin{equation}
\mathcal{I}_{sunset}(k,z_c;L) = \frac{L^4 \lambda^2}{4} \, \int_{\mathcal{D}} \frac{dz}{z^2} \, \frac{dw}{w^2} \int_{\mathbb{R}} \frac{d\vec{p}}{2\pi}\, G(z,w;p)G(z,w;|\vec{p}+\vec{k}|) \, K(z;k) \, K(w;k)~.
\end{equation}
For $\vec{k}=0$, the above integral can be performed analytically and we find:
\begin{equation}
\mathcal{I}_{sunset}(0,z_c;L) = \frac{L^4 \, \lambda^2}{4 \, z_c} \left( \frac{\pi^2-8}{8 \pi} \right)~.
\end{equation}
We were not able to perform the full integral analytically, however a numerical evaluation reveals the following small $z_c$ expansion:
\begin{equation}
\mathcal{I}_{sunset}(k,z_c;L) - \mathcal{I}_{sunset}(0,z_c;L)  =  \frac{L^4 \, \lambda^2}{2} \left(a_1 \, k \log k z_c + a_2 \, k + \ldots \right)~,
\end{equation}
with $a_1 \approx +0.261\ldots$ and $a_2 \approx +0.58\ldots$ The $(\varphi_{\vec{k}})^2$ piece of the late time (absolute value of the) wavefunction to order $\lambda^2$ is then:
\begin{equation}
\log|\Psi_{BD}[\varphi_{\vec{k}},\eta_c]| = \int \frac{d\vec{k}}{2\pi} \left( - \frac{k}{2} + \mathcal{I}_{sunset}(k,-i\eta_c;-i\ell) + \mathcal{I}_{tadpole}(k,-i\eta_c;-i\ell) \, \right) \, \varphi_{\vec{k}} \, \varphi_{-\vec{k}}~.
\end{equation}
Thus we see that at loop level there are logarithmic corrections to the $(\varphi_{\vec{k}})^2$ piece of the wavefunction. For the sunset diagram, the loop correction required no ultraviolet cancelation and so the logarithmic term present in the result is free of any potential scheme dependence. 

\section{Outlook}

In this paper we have explored the late time structure of $\Psi_{BD}$ in a de Sitter background, by computing its quantum corrections employing a perturbative framework heavily used in the AdS/CFT literature. We have identified several types of behavior, including the logarithmic growth in conformal time. Logarithmic growth commonly appears in the correlators computed in the in-in formalism. Furthermore, we have connected the late time properties of $\Psi_{BD}$ to certain anomalies and shifts in conformal weights of a CFT putatively dual to a bulk de Sitter theory containing the types of fields and interactions we studied. There are several interesting avenues left to explore. 
\begin{itemize}
\item {\it Graviton loops}: One would like to firmly establish the absence (or presence) of logarithmic growth for pieces of the wavefunction that depend on the metric only, both for a pure Einstein theory and more general theories of gravity, such as those with higher derivative terms. 

\item{\it Higher spin holography:} We found that cubic interactions for conformally coupled scalars lead to an additional local cubic piece of $\Psi_{BD}$ that was intricately related to a logarithmic phase. Such scalars are present in the higher spin Vasiliev theory, but the cubic scalar coupling is absent at the classical level \cite{Sezgin:2003pt,Petkou:2003zz}. At loop level, however, there may be a contribution to the cubic piece of $|\Psi_{BD}|$, which can be computed in the dual CFT. The presence of such additional local contributions may give interesting new contributions to $\Psi_{BD}$ for large field values. Similar considerations may also interesting for the alternate boundary condition dual to a $\Delta=1$ scalar operator in the CFT.

\item{\it Resummation:} We discussed a possible interpretation of the logarithmic growths as pieces of a series corresponding to a small shift in the conformal weight $\Delta$ of an operator in the dual CFT. For a massless scalar with $\phi^4$ self-interactions, we saw that such a shift would cause the dual operator to be marginally relevant, $\Delta<3$, rather marginally irrelevant. It would be interesting to relate this picture of resummation to other proposals involving dynamical renormalization group methods (see for example the review \cite{Seery:2010kh}).

\item {\it Stochastic inflation}: It would be interesting to relate our calculations/interpretations to the framework of stochastic inflation \cite{Starobinsky:1994bd} which proposes a non-perturbative approach for interacting fields in a fixed de Sitter background. Another approach to study strongly coupled (conformal) field theories in a fixed de Sitter background is using the AdS/CFT correspondence where AdS has a de Sitter boundary metric, on which the CFT resides (see for example \cite{Marolf:2010tg}).

\end{itemize}

\section*{Acknowledgements}

We would like to acknowledge useful discussions with Frederik Denef, Matt Dodelson, Daniel Green, Daniel Harlow, Sean Hartnoll, Paul McFadden, Anastasios Petkou, Edgar Shaghoulian and Julian Sonner. The research of DZF is supported in part by NSF grant PHY-0967299. Both DZF and TA are supported in part by the U.S. Department of Energy under cooperative research agreement DE-FG02-05ER41360. DA and GK are also partially funded by DOE grant DE-FG02-91ER40654.

\appendix

\section{A quantum mechanical toy model}\label{toy}

We consider a simple quantum mechanical toy model that captures some of the essence and mathematics of our (A)dS calculations. The Hamiltonian governing the system is given in the $\hat{x}$-eigenbasis by:
\begin{equation}
\hat{H} = -\frac{1}{2}\frac{d^2}{dx^2} + \frac{m^2 x^2}{2}  + \frac{\lambda}{6 \, t^2} \, x^3~, \quad\quad \hbar = 1~,
\end{equation}
where $x\in \mathbb{R}$ and the time parameter $t \in (-\infty,0)$ with $t \to 0$ as the infinitely late time limit of the system. The cubic interaction term is taken to be small and the dimensionless quantity $\alpha \equiv \lambda \, m^{-1/2}$ will serve as the small parameter in our perturbative analysis. The system has a scaling relation: $x \to  \nu^{1/2} x$, $t \to \nu t$, $m \to m/\nu$ and $\lambda \to \lambda/\nu^{1/2}$, which we could use to set $m=1$. Note that the cubic term is smaller than the quadratic term whenever:
\begin{equation}
\lambda \ll m^2 \langle x^2 \rangle^{-1/2} \, t^2~,
\end{equation}
The above Hamiltonian is unbounded from below, but this will be of no concern at the perturbative level. Moreover, if the state of interest is normalizable at a given time, the Hermiticity of the above Hamiltonian is enough to ensure that it will remain normalizeable for all times. The Schr\"{o}dinger equation governing the time evolution of a quantum state $\psi$ is given by:
\begin{equation}\label{schro}
i \, \partial_t \, \psi(x,t) = \hat{H} \, \psi(x,t)~.
\end{equation}
At $\lambda = 0$, we have that the ground state of the system is given by:
\begin{equation}
\psi_g(x,t) = \left(\frac{\pi}{m} \right)^{1/4} \, \exp \left( - \frac{i\,m}{2} \, t - \frac{m}{2} \, x^2 \right)~.
\end{equation}
The above state can be built from a Euclidean path integral with vanishing boundary conditions for $x(t)$ in the infinite Euclidean past $\tau \to \infty$, where $\tau \equiv - it$. For such a state we have that $\langle x^2 \rangle^{1/2} \sim 1/m^{1/2}$. 

We are interested in perturbations of the above wavefunction, i.e. solutions to the Schr\"{o}dinger equation that are continuously connected to $\psi_g$ in the limit $\lambda \to 0$. 

\subsection{Path integral perturbation theory}

The quantum states of interest can be constructed via a Euclidean path integral:
\begin{equation}
\psi_g^{(\lambda)}(\tilde{x},\tau_c) = \mathcal{N} \int \mathcal{D}x \, e^{-S_E(x)}~, 
\end{equation}
where $S_E$ is the Euclidean action governing the path integral:
\begin{equation}
S_E = \int d\tau \left( \frac{1}{2} \, \dot{x}^2 + \frac{1}{2}\, m^2 x^2 + \frac{\lambda}{6} \, \frac{x^3}{\tau^2}\right)~.
\end{equation}
As in the unperturbed case, the path integral is supplemented with the boundary conditions that $x(\tau) \to 0$ in the limit $\tau \to \infty$, and $x(\tau_c) = \tilde{x}$ (where $\tau_c>0$ is a late time cutoff). We consider a solution to the classical equations of motion $x_{cl}$ obeying the prescribed boundary conditions, supplemented by a quantum fluctuation $\delta x$. The path integral then splits as:
\begin{equation}
\psi_g^{(\lambda)}(\tilde{x},\tau_c) = e^{-S_E[x_{cl}]} \int \mathcal{D}\delta x \, e^{-S_E[\delta x]}~.
\end{equation}
Perturbatively, the solution can be expanded as $x_{cl} = x_0 + \lambda x_1 + \lambda^2 x_2 + \ldots$ We absorb the boundary dependence fully into the $x_0$ term. Thus we have:
\begin{equation}
x_0(\tau) = \tilde{x} \, e^{m(\tau_c-\tau)}~, \quad\quad x_1(\tau) = -\frac{1}{2} \int_{\tau_c}^\infty \frac{d\tau'}{(\tau')^2} \, x_0(\tau')^2 \, G(\tau,\tau')~,
\end{equation}
and so on. The `bulk-to-bulk' propagator $G(\tau,\tau')$ obeys:
\begin{equation}
\left( -\frac{d^2}{d\tau^2} + m^2 \right) \, G(\tau,\tau') = \delta(\tau-\tau')~.
\end{equation}
Explicitly:
\begin{equation}
G(\tau,\tau') = -\frac{1}{2m} \left( e^{2 m \tau_c}\, e^{-m(\tau+\tau')} - e^{m(\tau-\tau')} \right)~, \quad \tau < \tau'~,
\end{equation}
and similarly for $\tau>\tau'$. The classical action on such a solution is given by:
\begin{equation}
-S_E[x_{cl}] = \frac{1}{2}\, x_0 \dot{x}_{cl}|_{\tau = \tau_c}  - \frac{\lambda}{12} \int_{\tau_c}^{\infty} \frac{d\tau}{\tau^2} \, x_{cl}^3~.
\end{equation}
It captures the tree-level diagrams of the perturbative expansion. 

As a concrete example, at order $\lambda$, the cubic in $\tilde{x}$ contribution to the exponent of the (Euclidean) wavefunction is given by:
\begin{equation}
\frac{\lambda}{6} \, \tilde{x}^3 \, \int_{\tau_c}^{\infty} \frac{d\tau}{\tau^2} \, K(\tau_c,\tau)^3 =\frac{\lambda}{6} \, \tilde{x}^3 \,  \left(\frac{1}{{\tau_c}} \, +\, 3 e^{3 m {\tau_c}} \, m \, \text{Ei}(-3 m \, {\tau_c})\right)~,
\end{equation}
where we have defined the `bulk-to-boundary' propagator:
\begin{equation}
K(\tau_c,\tau') \equiv \lim_{\tau\to\tau_c} \partial_{\tau} G(\tau,\tau') = e^{m(\tau_c-\tau')}~.
\end{equation}
A late time expansion of the cubic correction yields:
\begin{equation}
\frac{\lambda }{6 {\tau_c}}+\frac{\lambda \, m}{2} \left(\gamma_E+{\log}(3 \, m \,{\tau_c}) \right)+\ldots
\end{equation}
We see that there are $1/\tau_c$ terms and $\log \tau_c$ that grow and eventually violate the perturbative assumption. To make contact with the ordinary Schr\"{o}dinger equation, we must analytically continue $\tau_c = -i t_c$. The $\sim 1/\tau_c$ term then becomes a contribution to the phase of the wavefunction and plays no role in its absolute value. On the other hand, the logarithmic term retains real part upon analytic continuation of the time and thus contributes to the absolute value of $\psi^{(\lambda)}_g(\tilde{x},t_c)$. Furthermore, for times $t \, m \sim e^{-1/\alpha}$ the cubic correction of the wavefunction becomes comparable to the $\lambda=0$ piece.

As another example, we can consider a diagram involving a loop, namely a tadpole diagram contributing a linear in $\tilde{x}$ piece to the exponent of the wavefunction. The correction is given by:
\begin{equation}
\frac{\lambda}{2} \, \int^\infty_{\tau_c} \frac{d\tau}{\tau^2} \, K(\tau_c,\tau) \, G(\tau,\tau)  = \frac{e^{m {\tau_c}}}{4} \, \left(3 \,e^{2 m {\tau_c}} \, {\text{Ei}}(-3 m {\tau_c})+{\text{Ei}}(-m {\tau_c})\right) ~.
\end{equation}
As for the cubic correction, a small $\tau_c$ expansion reveals logarithmic terms. The presence of a non-vanishing tadpole also implies that the vev of $\hat{x}$ is non-vanishing and in fact time dependent. A small $t$ expansion renders to order $\lambda$:
\begin{equation}
\langle \psi_g^{(\lambda)} | \hat{x} |  \psi_g^{(\lambda)} \rangle =  \frac{\lambda}{4\,m}  (\gamma_E+{\log}(-m t))+\ldots~
\end{equation}
Thus, for $\lambda > 0$ and $m > 0$ and at small enough $t$ the vev of $\hat{x}$ drifts to negative values where it will become sensitive to the unbounded part of the potential. 

In this fashion, using as the basic propagators of our perturbation theory $G$ and $K$, we can build the quantum corrections of the ground state $\psi_g(\tilde{x},t_c)$ at some time $t_c = -i\tau_c~$. In this simple example, one can explicitly check that the corrected wavefunction indeed solves the time dependent Schr\"{o}dinger equation (\ref{schro}) to the appropriate order in $\lambda$. 

\section{The bulk-to-bulk propagator}\label{GK} 

The Green's function for the massless scalar in Euclidean AdS$_{(d+1)}$ satisfies the partial differential equation:
\be\label{bulkprop}
\partial_\mu(\sqrt{g}g^{\mu\nu}\partial_\nu G(z,w)) = - \d^{(d+1)}(z-w)\,.
\ee
In this form the right side contains the naive $\d$-function, no $1/\sqrt{g}.$
The derivatives are taken with respect to the observation point $z^\m$ while $w^\m$ is the source point.   We will enforce the symmetry $G(z,w) = G(w,z).$  We really need the Green's function in momentum space:
\be
G(z,\vec{x}; w, \vec{y}) = \int \frac{d^d\vk}{(2\pi)^d} e^{i\vk\cdot(\vx -\vec{y})}
G(z,w,k)\,.
\ee
This satisfies  the second order ordinary differential equation:
\be \label{ode}
\left(\frac{L}{z}\right)^{(d-1)} \left( \pa_z^2  -\frac{(d-1)}{z}\pa_z-k^2 \right) G(z,w,k)= 0 \quad   z\ne w~.
\ee
First we choose a simple basis for the homogeneous modes of this equation.  The basis contains exponentially damped modes, called $\phi_2(z)$ below, as $z\to \infty$, and exponentially growing modes, called $\phi_1(z)$, obtained using the reflection symmetry $z \lra -z$ of the ODE.  For the two cases $D=2,\,4$ we write:
\bea \label{modes}
D=2 &&   \quad \phi_1(z) = e^{kz}  \qquad\quad\quad \phi_2(z) = e^{-kz}~,\\
D=4 &&   \quad \phi_1(z) = (1-kz)e^{kz}  \qquad\quad \phi_2(z) = (1+kz)e^{-kz}~.
\eea
In \cite{Gubser:2002zh} the bulk Green's function was constructed using a different choice of basis modes. The Green's function for a second order ODE is commonly treated in texts on differential equations, and we have used Ch. 9 of \cite{birkhoffRota}. The Green's function is the product of modes in the two sectors $z<w$ and $z>w$:
\bea
G(z,w,k) &=&  A \phi_1(z)\phi_2(w) +  c \phi_2(z)\phi_2(w)   \quad\quad   z<w~,\\
&=&   B \phi_2(z)\phi_1(w)  +c \phi_2(z)\phi_2(w)   \quad\quad   z>w~.
\eea
Note that we always choose the exponentially damped mode for the larger of the two variables.  The coefficients $A,\,B$ are determined by the following conditions at the ``diagonal" point $z=w$:
\begin{itemize}
\item  \text{$G(z,w,k)$ is continuous at $z=w~$,}\\
\item  \text{the first derivative $\pa_zG(z,w,k)$ decreases by $(z/L)^{(d-1)}$  as $z$ increases through $z=w$~.}
\end{itemize}
For the ODE in the form (\ref{ode}),  \cite{birkhoffRota} specifies that the jump is the reciprocal of the leading coefficient as we have written.  These conditions uniquely determine $A,\,B$, but not $c$ since it multiplies a product of modes that is smooth across the diagonal.  
\begin{itemize}
\item \text {$c$ is determined by enforcing the Dirichlet boundary condition  $G(z=z_c, w,k) = 0$ at the cutoff.}
\end{itemize}
It is easy to see that these conditions completely determine the Green's function. 
 In two bulk dimensions we have the expression:
\bea  \lab{green2}
G(z,w,k) &=& \frac{1}{2k} \left[    \phi_1(z)\phi_2(w) -\frac{\phi_1(z_c)\phi_2(z)\phi_2(w)}{\phi_2(z_c)}  \right]  \quad\quad   z<w~,\\
&=& \frac{1}{2k} \left[  \phi_2(z)\phi_1(w)- \frac{\phi_1(z_c)\phi_2(z)\phi_2(w)}{\phi_2(z_c)}\right]   \quad\quad   z>w\,.
\eea
and in four bulk dimensions we have:  
\bea\lab{green4}
G(z,w,k) &=&- \frac{1}{2k^3L^2} \left[\phi_1(z)\phi_2(w) - \frac{\phi_1(z_c)\phi_2(z)\phi_2(w)}{\phi_2(z_c)}\right]   \quad\quad   z<w~,\\
&=&- \frac{1}{2k^3L^2} \left[\phi_2(z)\phi_1(w) - \frac{\phi_1(z_c)\phi_2(z)\phi_2(w)}{\phi_2(z_c)}\right]   \quad\quad   z>w~.
\eea

One very good check of these results comes enforcing the correct relation between the bulk-to-bulk and bulk-to-boundary propagators. This follows from the application of Green's formula to the boundary value problem:
\be \lab{bvp}
\pa_\m \sqrt{g}g^{\m\n}\pa_\n \, \phi(z,\vx) = 0~, \qquad  \phi(z_c,\vx) = \varphi(\vx)\,.
\ee
Green's formula reads  (note $z_c=w_c$)
\bea 
&&\int_{w_c}^\infty dw \int d^d\vec{y} \sqrt{g(w)} \left(\phi(w)\Box_w  G(w,z) -G(w,z)\Box_w\phi(w)\right) \nonumber\\
&&\qquad= \int_{w_c}^\infty dw \int d^d\vec{y} \,\, \pa_\m( \sqrt{g(w)}g^{\m\n} (\phi(w)\pa_\n G(w,z)- G(w,z) \pa_w \phi(w)))~,\lab{grfo}
\eea
and thus:
\be
-\phi(z,\vx)=-\int d^d\vec{y} \left(\frac{L}{w_c}\right)^{(d-1)} \pa_w G(w=w_c,\vec{y};z,\vec{x}) \varphi(\vec{y})\,.
\ee
To reach the last expression we use \reef{bulkprop} and the fact that the PDE \reef{bvp} has no bulk source, and we evaluate the second line at the boundary $w=w_c$ where the Dirichlet Green's function vanishes.  The main point is that the bulk-to-boundary propagator $K(z,\vx)$ is the properly normalized radial derivative of the bulk-to-bulk Green's function;  the specific relation is
\be \lab{btb-btb}
K(z,\vx-\vec{y}) =  \sqrt{g(w_c)}g^{ww} \pa_w G(w_c,\vec{y};z,\vx) \,.
\ee
After Fourier transformation, the last expression of (\ref{grfo}) exactly reproduces the solution of the linearized solution of the $k$ space EOM (\ref{adseom}) for both $D=2,\,4$.

\end{document}